\documentstyle[pre,aps,epsf]{revtex}

\begin{document}
\draft

\title{Dynamic Phase Transition, Universality, and Finite-size Scaling in the 
Two-dimensional Kinetic Ising Model in an Oscillating Field}
\author{G. Korniss,$^1$\footnote{Permanent address: 
Department of Physics, Applied Physics, and Astronomy, 
Rensselaer Polytechnic Institute, Troy, NY 12180-3590.}
C.~J. White,$^{1,2}$ P.~A. Rikvold,$^{1,2}$ and 
M.~A. Novotny$^1$}
\address{$^1$School of Computational Science and Information Technology, 
Florida State University, Tallahassee, Florida 32306-4120 \\
$^2$Center for Materials Research and Technology and Department of 
Physics, Florida State University, Tallahassee, Florida 32306-4350}

\date{\today}
\maketitle
\begin{abstract}
We study the two-dimensional kinetic Ising model below its equilibrium 
critical temperature, subject to a square-wave oscillating external field.
We focus on the multi-droplet regime where the metastable phase decays
through nucleation and growth of {\em many} droplets of the stable phase. 
At a critical frequency, the system undergoes a genuine non-equilibrium phase 
transition, in which the symmetry-broken phase corresponds to an asymmetric 
stationary limit cycle for the time-dependent magnetization. 
We investigate the universal aspects of this dynamic phase transition at 
various temperatures and field amplitudes via large-scale Monte Carlo 
simulations, employing finite-size scaling techniques adopted from 
equilibrium critical phenomena. The critical exponents, the fixed-point value
of the fourth-order cumulant, and the critical order-parameter distribution
all are consistent with the universality class of the two-dimensional 
{\em equilibrium} Ising model. We also study the cross-over from the 
multi-droplet to the strong-field regime, where the transition disappears.
\end{abstract}
\pacs{PACS numbers: 
64.60.Ht, 
75.10.Hk, 
64.60.Qb, 
05.40.-a  
}

\section{Introduction}

Metastability and hysteresis are widespread phenomena in nature.
Ferromagnets are common systems that exhibit these behaviors
\cite{EWIN1881,WARB1881,EWIN1882,STEI1892,AHARONI}, but there are also
numerous other examples ranging from ferroelectrics \cite{BEALE,RAO91}
to electrochemical adsorbate layers \cite{SMEL,MITCHELL00} 
to liquid crystals \cite{CHENG96}. 
A simple model for many of these real systems is the kinetic Ising 
model (in either the spin or the lattice-gas representation). For example,
it has been shown to be appropriate for describing magnetization dynamics in 
highly anisotropic single-domain nanoparticles and uniaxial thin films 
\cite{HE,JIANG,SUEN,group2}. 
\begin{figure}
\center
\vspace*{-2.5cm}
\epsfxsize=18cm \epsfysize=18cm \epsfbox{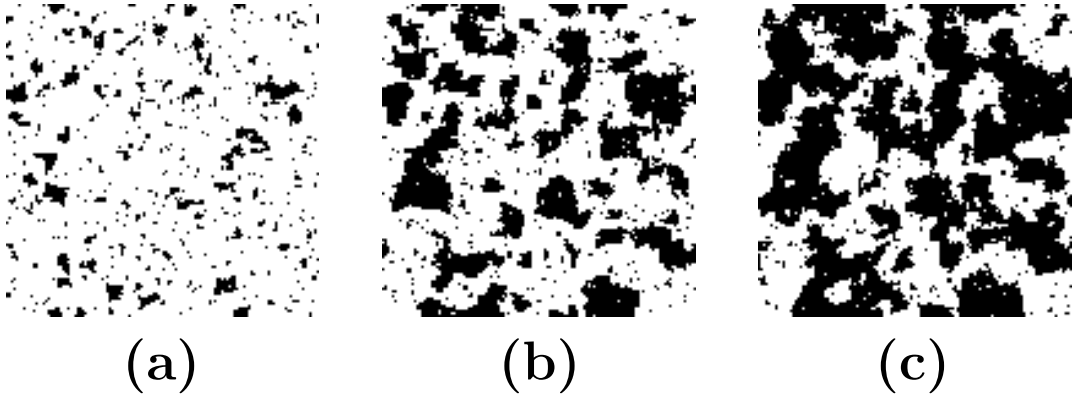}
\vspace*{-10cm} 
\caption{Metastable decay in the multi-droplet regime at $T$$=$$0.8T_{\rm c}$ 
for an $L$$=$$128$ square-lattice kinetic Ising system evolving under Glauber 
dynamics. The system is initialized with all spins 
$s_i$$=$$1$, and an applied field, $H$$=$$-0.3J$, is set at $t$$=$$0$. 
Snapshots of the spin configurations are given at
(a) $t$$=$$30$ Monte Carlo steps per spin (MCSS), 
(b) $t$$=$$60$ MCSS,
(c) $t$$=$$74$ MCSS.
The metastable lifetime (the average first-passage time to zero magnetization)
at this temperature and field is $\langle\tau\rangle$$=$$74.5$ MCSS. 
Stable $s$$=$$-1$ (metastable $s$$=$$+1$) spins are represented by black 
(white).}
\label{decay_conf}
\end{figure}

The system response to a single reversal of the ``external field'' is fairly 
well understood \cite{switch}. In sufficiently large systems 
below the equilibrium critical temperature, $T_c$,
the order parameter changes its value through the nucleation and 
growth of {\em many} droplets, inside which it has the equilibrium value 
consistent with the value of the applied field, as shown in 
Fig.~\ref{decay_conf}.  
This is the multi-droplet regime of phase transformation 
\cite{switch,phase_transformation}. The well-known
Avrami's law \cite{Avrami} describes this process of homogeneous nucleation 
followed by growth quite accurately up to the time when
the growing droplets coalesce and the stable phase becomes the majority
phase \cite{RAMOS99}. 
The intrinsic time scale of the system is given by the 
metastable lifetime, $\langle\tau\rangle$, which is defined as the 
average first-passage time to zero magnetization. It is a measure of the
time it takes for the system to escape from the 
metastable region of the free-energy landscape.
In this paper we will use the magnetic language in which the order parameter
is the magnetization, $m$, and its conjugate field is the external magnetic 
field, $H$. Analogous interpretations, e.g., using the terms
polarization and electric field for ferroelectric systems \cite{BEALE,RAO91}, 
and coverage and chemical potential for adsorption problems 
\cite{SMEL,MITCHELL00}, are straightforward. 

It is natural next to ask, ``what is the response to an oscillating external 
field?'' The hysteretic behavior in ferromagnets has attracted 
significant experimental interest, mainly focused on the characteristic 
behavior of the hysteresis loop and its area. Its dependence on the field 
amplitude and frequency has been intensively studied and its scaling behavior 
(power law versus logarithmic) is still under investigation, both 
experimentally \cite{HE,JIANG,SUEN} and theoretically 
\cite{JUNG90,RAO90,TOME90,Mendes91,Zimmer93,LO90,SIDES98a,SIDES98b,SIDES99}. 
For the kinetic Ising ferromagnet in two dimensions it has been recently shown 
\cite{SIDES98a,SIDES98b,SIDES99} that the true behavior is in fact a crossover,
approaching a logarithmic frequency dependence only for extremely low 
frequencies. 

\begin{figure}
\center
\epsfxsize=6cm \epsfysize=6cm \epsfbox{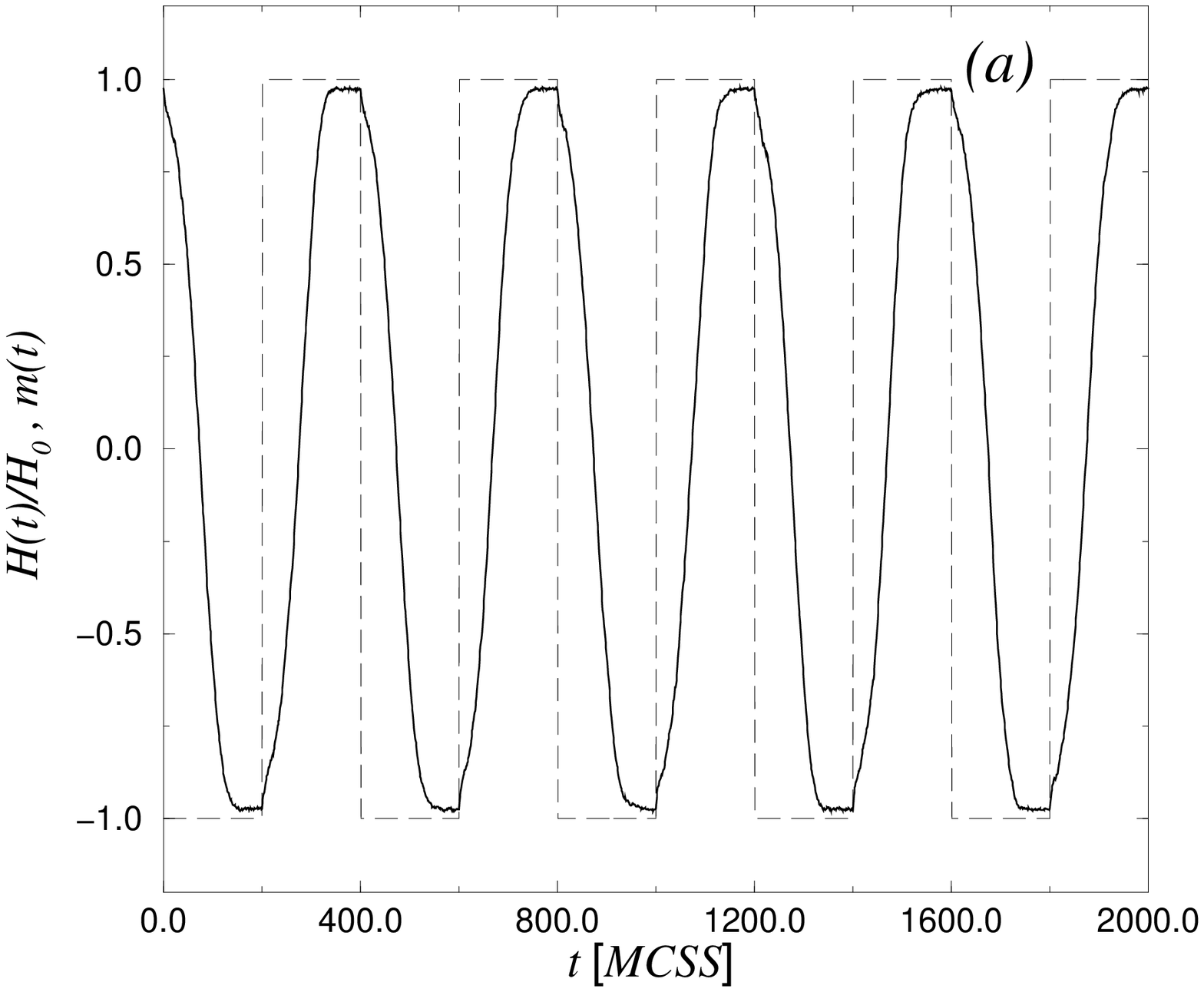}
\epsfxsize=6cm \epsfysize=6cm \epsfbox{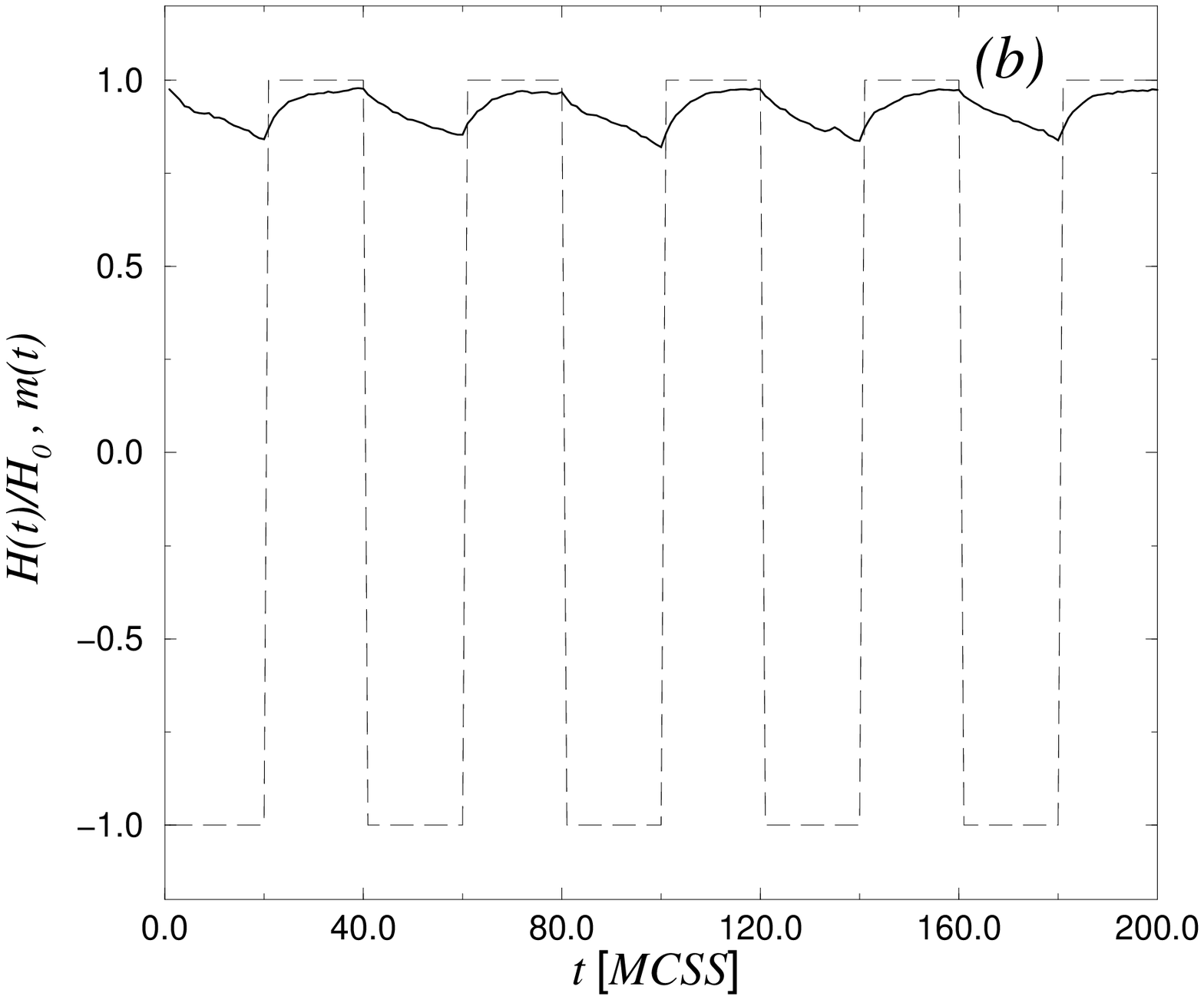}
\caption{Monte Carlo magnetization time series (solid lines) in the presence 
of a square-wave external field (dashed lines) for an $L$$=$$128$ system at 
$T$$=$$0.8T_{\rm c}$ and field amplitude $H_{0}$$=$$0.3J$. ($T_{\rm c}$
is the two-dimensional equilibrium Ising critical temperature.)
The metastable 
lifetime at this temperature and field is $\langle\tau\rangle$$=$$74.5$ MCSS. 
(a) Dynamically disordered phase at dimensionless half-period 
$\Theta$$\equiv$$t_{1/2}/\langle\tau\rangle$$=$ $2.7$. 
(b) Dynamically ordered phase at $\Theta$$=$$0.27$.}
\label{m_series}
\end{figure}

An important aspect of hysteresis in bistable systems, which is the focus
of the present paper, is the dynamic competition between the two time scales 
in the system: the half-period of the external field  $t_{1/2}$ (proportional 
to the inverse of the driving frequency) and the metastable lifetime 
$\langle\tau\rangle$.
For low frequencies, a complete decay of the metastable phase almost always
occurs in each half-period, just like it does after a single field reversal. 
Consequently, the time-dependent magnetization reaches a limit cycle 
which is symmetric about zero [Fig.~\ref{m_series}(a)]. 
For high frequencies, however, the system does not have enough time
to switch during one half-period, and the symmetry of the hysteresis loop is 
broken. The magnetization then reaches an asymmetric limit cycle 
[Fig.~\ref{m_series}(b)]. 
Avrami's law \cite{Avrami,RAMOS99} is a good approximation when the majority 
of the droplets do not overlap. Thus, it can be employed to estimate the 
time-dependent magnetization and the dynamic order parameter 
(period-averaged magnetization) in the low-frequency (see the Appendix) and 
in the high-frequency \cite{SIDES99} limits. However, it cannot describe the 
``critical regime'' where $t_{1/2}$ becomes comparable to $\langle\tau\rangle$,
and which is dominated by coalescing droplets. 
\noindent
This symmetry breaking between the symmetric and asymmetric limit cycles 
has been the subject of intensive research over the last decade. 
It was first observed during numerical integration of a 
mean-field equation of motion for the magnetization of a ferromagnet in an 
oscillating field \cite{TOME90,Mendes91}. Since then, it has been observed
and studied in numerous Monte Carlo (MC) simulations of kinetic Ising systems
\cite{LO90,SIDES99,SIDES98,ACHA95,ACHA97C,ACHA97D,ACHA98,BUEN00}, as well as 
in further mean-field studies \cite{Zimmer93,ACHA95,ACHA97D,ACHA98,BUEN98}. 
It may also have been 
experimentally observed in ultrathin films of Co on Cu(001) \cite{JIANG}.
The results of these studies suggest that this symmetry breaking corresponds 
to a genuine continuous non-equilibrium phase transition. For recent reviews 
see Refs. \cite{ACHA94,CHAK99}. Associated with the transition
is a divergent time scale (critical slowing down) \cite{ACHA97D} and, for 
spatially extended systems, a divergent correlation length 
\cite{SIDES99,SIDES98}. 
Estimates for the critical exponents and the universality class of the 
transition have recently become available \cite{SIDES99,SIDES98,UGA99}
after the successful application of finite-size scaling techniques borrowed 
from equilibrium critical phenomena \cite{FISH72,BIND92,BIND90,LANDAU76}.

The purpose of the present paper is to extend preliminary results \cite{UGA99}
and to provide more accurate estimates of the exponents for two-dimensional 
kinetic Ising systems in a square-wave oscillating field.
The use of the square-wave field tests the universality of the dynamic 
phase transition (DPT) \cite{SIDES99,SIDES98}, and it also significantly 
increases computational speed, compared to the more commonly used sinusoidal 
field. We further explore the universal aspects of the transition by varying 
the temperature and field amplitude within the 
multi-droplet regime, and we study the cross-over to the strong-field regime
where the transition disappears.
In obtaining our results, we rely on dynamic MC simulations.
Computational methods are always helpful, especially when theoretical
ideas are largely missing. There are cases, however, when even the use of 
standard equilibrium techniques, such as finite-size scaling requires some 
insight and building analogies between equilibrium and non-equilibrium 
systems \cite{SIDES99,SIDES98}. This is the case for our present study. 
No effective ``Hamiltonian'' was known before the completion of this work for 
the dynamic order-parameter (in the 
coarse-grained sense), from which the long-distance behavior of the model could
be derived. This is a typical difficulty when dealing with systems far from 
equilibrium \cite{DDS,MARRO_DICKMAN}. Recently, however, a coarse-grained 
Hamiltonian has been derived \cite{Fuji} for the dynamic order-parameter, 
supporting our results for the DPT.
Similar to the previous work for sinusoidly oscillating fields 
\cite{SIDES99,SIDES98}, we perform large-scale simulations and 
finite-size scaling to investigate the universal properties of the DPT.

The remainder of the paper is organized as follows. 
In Sec.~II we define the model 
and the observables of interest.
Section~III contains the Monte Carlo results and 
analyses. Conclusions and outlook are given in Sec.~IV.

\section{Model and Relevant Observables}

To model spatially extended bistable systems in two dimensions, we study a
nearest-neighbor kinetic Ising ferromagnet on a $L$$\times$$L$ square lattice 
with periodic boundary conditions. The model is defined by the Hamiltonian 
\begin{equation}
\label{eq:Hamil}
{\cal H } = -J \sum_{ {\langle ij \rangle}} {s_{i}s_{j}}
               - H(t) \sum_{i} {s_{i}} \;,
\end{equation}
where $s_{i}$$=$$\pm 1$ is the state of the $i$th spin, 
$J > 0$ is the ferromagnetic interaction, 
$\sum_{ {\langle ij \rangle} }$ runs over all nearest-neighbor pairs,
$\sum_{i}$ runs over all $L^{2}$ lattice sites, and $H(t)$ is an oscillating, 
spatially uniform applied field. The magnetization per site,
\begin{equation}
\label{eq:m(t)} 
{m(t)} = \frac{1}{L^2} \sum_{i=1}^{L^2} {s_{i}(t)} \;,
\end{equation}
is the density conjugate to $H(t)$.
The temperature $T$ is fixed below its zero-field critical value 
$T_{\rm c}$ ($J/k_{\rm B} T_{\rm c}$$=$$ \ln(1+\sqrt 2)/2$ 
\cite{Onsager}, where $k_{\rm B}$ is Boltzmann's constant), so that the 
magnetization for 
$H$$=$$0$ has two degenerate spontaneous equilibrium values, 
$\pm m_{\rm sp}(T)$. For nonzero fields the equilibrium magnetization has
the same sign as $H$, while for $H$  not too strong, the opposite 
magnetization direction is {\it metastable\/} and decays slowly towards 
equilibrium with time, as described in Sec. I.

The dynamic used in this study, as well as in Refs.~\cite{SIDES99,SIDES98}, 
is the Glauber single-spin-flip MC algorithm with updates at randomly chosen 
sites. Note that the random sequential update scheme corresponds to 
independent Poisson arrivals for the update attempts (discrete events) at each
site. Thus, the arrival pattern is strongly asynchronous.
The time unit is one MC step per spin (MCSS). Each attempted spin flip 
from ${s_{i}}$ to ${-s_{i}}$ is accepted with probability 
\begin{equation}
\label{eq:Glauber}
W(s_{i} \rightarrow -s_{i}) =
\frac{ \exp(- \beta \Delta E_{i})}{1 + \exp(- \beta \Delta E_{i})} \; .
\end{equation}
Here $\Delta E_{i}$ is the energy change resulting from acceptance, 
and $\beta \! = \! 1/k_{\rm B}T$. 
For the largest system studied ($L$$=$$512$) we used a scalable massively
parallel implementation of the algorithm for this {\em asynchronous} 
dynamics \cite{Luba,PAR_JCP,PAR_PRL,UGA00}.
The parallel discrete-event scheme ensures that the underlying dynamic is not 
changed (that is, the update attempts are identical, independent Poisson 
arrivals at each site), while a substantial amount of parallelism is exploited.
The parallel implementation \cite{PAR_JCP} was carried out on a Cray T3E,
employing up to 256 processing elements.

The dynamic order parameter is the period-averaged magnetization \cite{TOME90},
\begin{equation}
\label{eq:Qeq}
Q = \frac{1}{2 t_{1/2}} \oint m(t) dt \;,
\end{equation}
where $t_{1/2}$ is the half-period of the oscillating field, and 
the beginning of the period is chosen at a time when $H(t)$ 
changes sign. Although the phase of the field does not influence the 
results reported in this paper, the choice made here is convenient in studies
of the hysteresis loop-area distributions and consistent with 
Refs.~\cite{SIDES98a,SIDES98b,SIDES99,SIDES98}.
Analogously we also define the local order parameter
\begin{equation}
\label{eq:Qlocaleq}
Q_{i} = \frac{1}{2 t_{1/2}} \oint s_{i}(t) dt \;,
\end{equation}
which is the period-averaged spin at site $i$. For 
slowly varying fields the probability distribution of $Q$ 
is sharply peaked at zero \cite{SIDES99,SIDES98}. 
We shall refer to this as the {\it dynamically disordered phase\/}. It is 
illustrated by the evolution of the magnetization in Fig.~\ref{m_series}(a) and
by the $Q$$\approx$$0$ time series in Fig.~\ref{Q_series}. 
For rapidly oscillating fields the distribution of $Q$
becomes bimodal with two sharp peaks near $\pm m_{\rm sp}(T)$, 
corresponding to the broken symmetry of the hysteresis loops 
\cite{SIDES99,SIDES98}. 
We shall refer to this as the {\it dynamically ordered phase}. It is 
illustrated in Fig.~\ref{m_series}(b) and by the 
$Q$$\approx$$m_{\rm sp}$$\sim$${\cal O}(1)$ time series in Fig.~\ref{Q_series}.
Near the DPT we use finite-size scaling analysis of 
MC data to estimate the critical exponents that characterize the transition. 
We also keep track of the normalized period-averaged internal energy (in units 
of $J$) \cite{ACHA97D},
\begin{equation}
E=-\frac{1}{2 t_{1/2}} 
\oint \frac{1}{L^2}\sum_{ {\langle ij \rangle}} {s_{i}(t)s_{j}(t)} dt \;,
\label{energy_def}
\end{equation}
since it also exhibits important characteristics of the DPT.

Previous studies of the DPT have used an applied field
which varies sinusoidally in time. While sinusoidal or 
linear saw-tooth fields are the most common in experiments and are 
necessary to obtain a vanishing loop area in the low-frequency 
limit \cite{SIDES98a,SIDES98b,SIDES99}, 
the wave form of the field should not affect universal 
aspects of the DPT. This should be so because the transition essentially 
depends on the competition between two time scales: the half-period $t_{1/2}$
of the applied field, and the average time it takes the system to 
leave the metastable region near one of its two degenerate 
zero-field equilibria when a field of magnitude $H_0$ and sign 
opposite to the magnetization is applied. 
This {\it metastable lifetime\/}\index{metastable lifetime}, 
$\langle \tau(T,H_0) \rangle$, 
is estimated as the average 
first-passage time\index{first-passage time} to zero magnetization. 
In the present paper we use a {\it square-wave\/} field of amplitude $H_0$.
This has significant computational advantages over the sinusoidal 
field variation since we can use two look-up tables to determine the 
acceptance probabilities: one for $H \!=\! + H_0$ and one for $H \!=\! - H_0$.

In terms of the dimensionless half-period, 
\begin{equation}
\label{eq:Theta}
\Theta = t_{1/2} \left/ \langle \tau(T,H_0) \rangle \right. \;,
\end{equation}
the DPT should occur at a critical value $\Theta_{\rm c}$ of order unity. 
Although $\Theta$ can be changed by varying either $t_{1/2}$, $H_0$, or $T$, 
in a first approximation we expect $\Theta_{\rm c}$ to depend only weakly 
on $H_0$ and $T$. This expectation will be confirmed in Sec. IV 
by simulations carried out at several values of $H_0$ and $T$ for different 
system sizes. 

In many studies of the DPT the transition has been approached 
by changing $H_0$ or $T$ \cite{LO90,ACHA95,ACHA97C,ACHA98}. 
While this is correct in principle, $\langle \tau(T,H_0) \rangle$ depends 
strongly and nonlinearly on its arguments \cite{switch}. We 
therefore prefer changing $t_{1/2}$ at constant $H_0$ and $T$ 
\cite{SIDES99,SIDES98}, 
as this gives more precise control over the distance from the 
transition. 

We focus on systems which are not only larger than the critical droplet, but
also significantly larger than the typical droplet separation \cite{switch}.
In this regime {\em many} supercritical droplets form and contribute to the 
decay of the metastable phase (the KJMA or Avrami theory for homogeneous 
nucleation \cite{Avrami,RAMOS99}), as seen in Fig.~\ref{decay_conf}.
This is the only regime where the DPT is expected to exist.
For small systems one observes subtle finite-size effects, not related to the 
DPT but rather to the {\em stochastic} single-droplet decay mode 
\cite{switch,tric}. In the single-droplet regime, subject to a 
periodic applied field, the system exhibits stochastic resonance 
\cite{SIDES98b}.

\section{Simulation results}
\subsection{Signs of the dynamic phase transition}

We performed extensive simulations and finite-size scaling analysis of the 
data on square lattices with $L$ between $64$ and $512$ at $T$$=$$0.8T_c$ and 
$H_0$$=$$0.3J$. We also
investigated the universality of the DPT within the multi-droplet regime
for various fields and temperatures below the equilibrium critical temperature,
using smaller systems with $L$ from $64$ to $128$. 
Typical runs near the DPT consist of $2$$\times$$10^5$ full periods. 
For example, at $T$$=$$0.8T_c$ and $H_0$$=$$0.3J$, where the critical 
half-period of the field is about $70$ MCSS, this corresponds to 
$2.8$$\times$$10^7$ MCSS.
Away from the transition point, an order of magnitude shorter runs were 
sufficient to obtain high-quality statistics.

\begin{figure}
\center
\epsfxsize=6cm \epsfysize=6cm \epsfbox{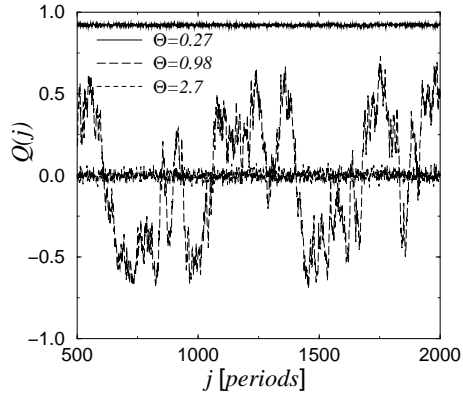}
\caption{Time series of the order parameter $Q$ 
at $T\!=\!0.8T_{\rm c}$ and $H_0\!=\!0.3J$ for $L\!=\!128$. 
Horizontal trace near $Q \!=\! +1$: 
$\Theta\!=\!0.27 < \Theta_{\rm c}$ 
[dynamically ordered phase, corresponding to Fig.~\protect\ref{m_series}(b)]. 
Strongly fluctuating trace: 
$\Theta\!=\!0.98 \! \approx \! \Theta_{\rm c}$ 
(near the DPT). 
Horizontal trace near $Q \!=\! 0$: 
$\Theta\!=\!2.7 > \Theta_{\rm c}$ 
[dynamically disordered phase, corresponding to 
Fig.~\protect\ref{m_series}(a)].}
\label{Q_series}
\end{figure}

The system was initialized with all spins up and the square-wave external 
field started with the half-period in which $H$$=$$-H_0$. After some relaxation
the system magnetization would reach a limit cycle [Fig.~\ref{m_series}] 
(except for thermal fluctuations). In other words, $Q$ [Eq.~(\ref{eq:Qeq})] 
(together with other period-averaged quantities) becomes a {\em stationary}
stochastic process [Fig.~\ref{Q_series}]. 
We discarded the first $1000$ periods of the time series to exclude 
transients from the stationary-state averages.

For large half-periods ($\Theta\gg\Theta_{\rm c}$) the magnetization switches 
every half-period [Fig.~\ref{m_series}(a)] and $Q$$\approx$$0$,
while for small half-periods ($\Theta\ll\Theta_{\rm c}$) 
the magnetization does not have time to switch during a single half-period 
[Fig.~\ref{m_series}(b)], 
resulting in $|Q|$$\approx$$m_{\rm sp}$, as can be seen from the time series 
in Fig.~\ref{Q_series}. 
The transition between the high- and low-frequency
regimes is characterized by large fluctuations in $Q$ near 
$\Theta_{\rm c}$ [Fig.~\ref{Q_series}].
\begin{figure}
\center
\vspace*{-2.5cm}
\epsfxsize=18cm \epsfysize=18cm \epsfbox{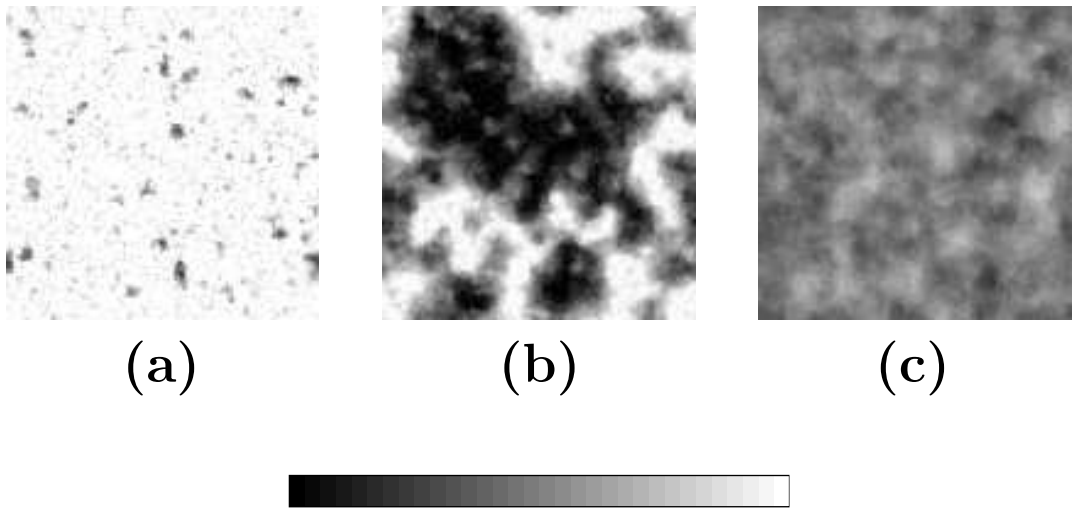}
\vspace*{-8cm} 
\caption{Configurations of the local order parameter $\{Q_i\}$ 
at $T$$=$$0.8T_{\rm c}$ and $H_0$$=$$0.3J$ for $L$$=$$128$. 
(a) $\Theta\!=\!0.27 < \Theta_{\rm c}$ (dynamically ordered phase). 
(b) $\Theta\!=\!0.98 \! \approx \! \Theta_{\rm c}$
(near the DPT).
(c) $\Theta\!=\!2.7 > \Theta_{\rm c}$ 
(dynamically disordered phase).
On the gray-scale black (white) corresponds to $-1$ ($+1$).}
\label{local_conf}
\end{figure}

To illustrate the spatial aspects of the transition 
we also show configurations of the local order parameter $\{Q_i\}$ 
in Fig.~\ref{local_conf}. Below $\Theta_{\rm c}$ [Fig.~\ref{local_conf}(a)] 
the majority of spins spend 
most of their time in the $+1$ state, i.e., in the metastable phase during the 
first half-period, and in the stable equilibrium phase during the second 
half-period (except for equilibrium fluctuations). 
Thus, most of the $Q_i$$\approx$$+1$. 
Droplets of $s_i$$=$$-1$ that nucleate during the negative half-period and 
then decay back to $+1$ during the positive half-period show up as 
roughly circular gray spots in the figure. Since the spins near the 
center of such a droplet become negative first and revert to positive last, 
these spots appear darkest in the middle. 
Also, for not too large lattices one occasionally observes the full reversal 
of an ordered configuration $\{Q_i\}$$\rightarrow$$\{-Q_i\}$, typical of 
finite, spatially extended systems undergoing symmetry breaking.
Above $\Theta_{\rm c}$ [Fig.~\ref{local_conf}(c)] the system follows the field
in every half-period (with some phase lag) and $Q_i$$\approx$$0$ 
at all sites $i$. 
Near $\Theta_{\rm c}$ [Fig.~\ref{local_conf}(b)] there are large clusters of
both $Q_i$$\approx$$+1$ and $Q_i$$\approx$$-1$ separated by 
``interfaces'' where $Q_i$$\approx$$0$. These large-scale structures remain 
reasonably stationary over several periods.

\begin{figure}
\center
\epsfxsize=5.5cm \epsfysize=5.5cm \epsfbox{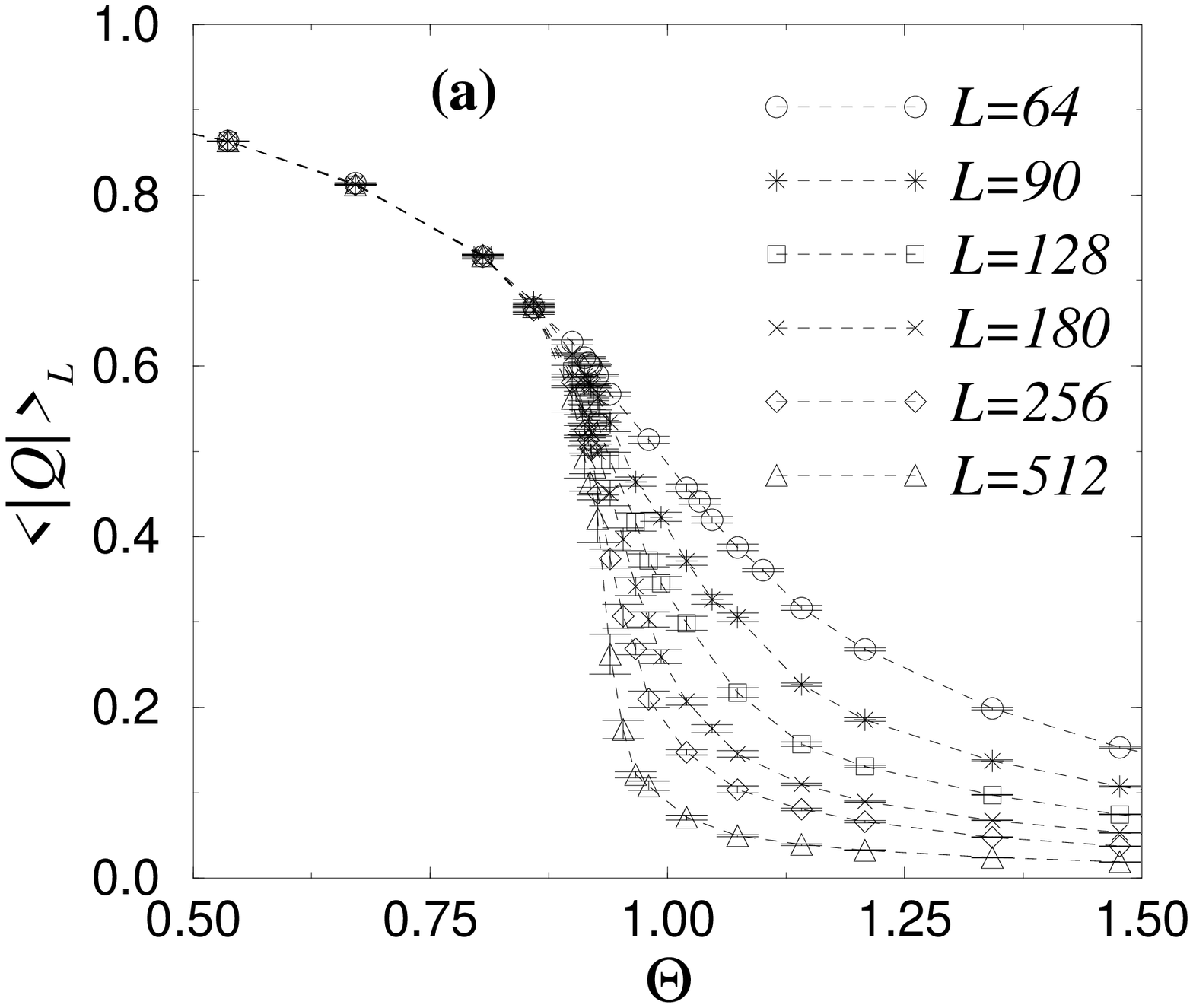}
\epsfxsize=5.5cm \epsfysize=5.5cm \epsfbox{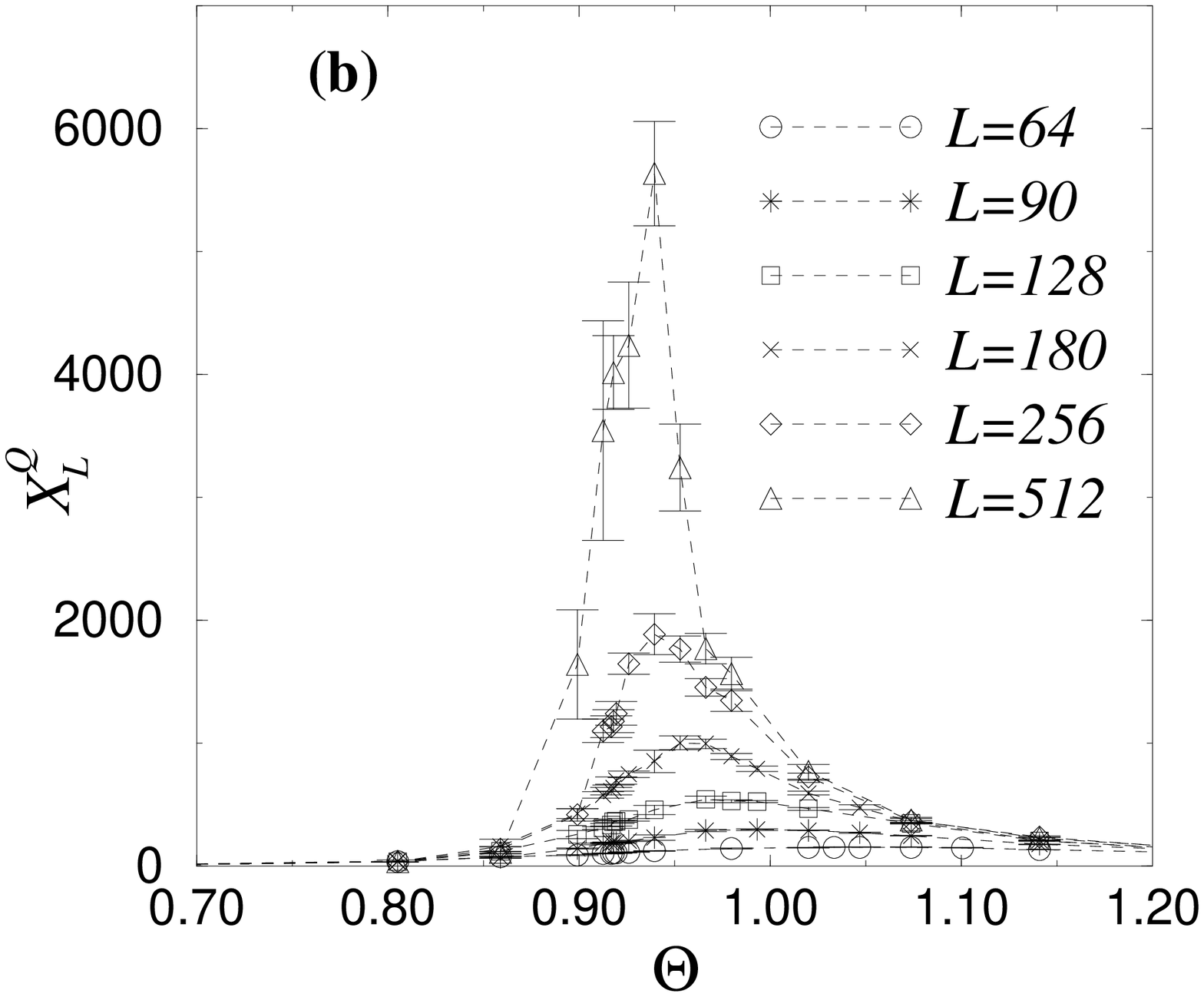}
\epsfxsize=5.5cm \epsfysize=5.5cm \epsfbox{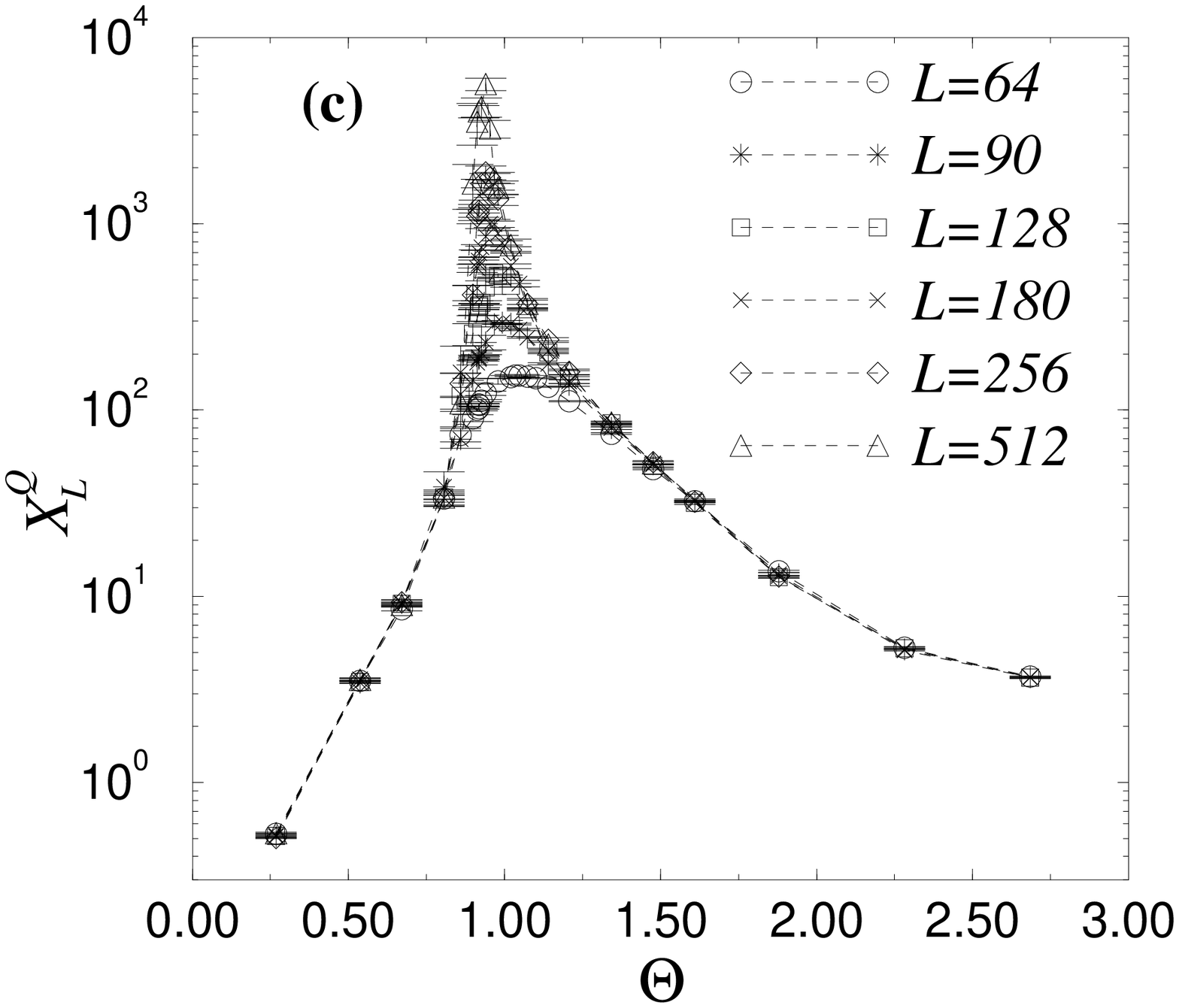}
\caption{Finite-size behavior of the order parameter at $T$$=$$0.8T_{\rm c}$
and $H_0$$=$$0.3J$ for various system sizes.
(a) The order parameter $\langle|Q|\rangle_{L}$.
(b) The scaled variance of the order parameter, $X^Q_L$, as defined in 
Eq. (\protect\ref{eq:XQ}).
(c) Same as (b) on lin-log scale to provide an enhanced view of the peaks for 
smaller systems.}
\label{Q_raw}
\end{figure}

For finite systems in the dynamically 
ordered phase the probability density of $Q$ becomes 
bimodal. Thus, to capture symmetry breaking, one has to measure the average
norm of Q as the order parameter, i.e., $\langle |Q| \rangle$ \cite{BIND92}. 
Figure~\ref{Q_raw}(a) shows that this order parameter is of order
unity for $\Theta<\Theta_{\rm c}$ and vanishes for $\Theta>\Theta_{\rm c}$, 
except for finite-size effects. 
To characterize and quantify this transition in terms of critical exponents
we employ the well-known technique of finite-size 
scaling \cite{FISH72,BIND92,BIND90,LANDAU76}.
The quantity analogous to the susceptibility is the scaled variance
of the dynamic order parameter \cite{SIDES99,SIDES98}, 
\begin{equation}
X^{Q}_{L}=L^2 \left(\langle Q^2 \rangle_L -\langle |Q|\rangle_L ^2 \right) \;.
\label{eq:XQ}
\end{equation}
Note that for our system the field conjugate to $Q$ and a corresponding 
fluctuation-dissipation theorem are not known, hence we cannot measure the 
susceptibility directly.
For finite systems $X_L$ has a characteristic peak near $\Theta_{\rm c}$ 
[see Fig.~\ref{Q_raw}(b)] 
which increases in height with increasing $L$, while no finite-size effects
can be observed for $\Theta\ll\Theta_{\rm c}$ 
and $\Theta\gg\Theta_{\rm c}$. This implies
the existence of a divergent length scale, possibly the 
correlation length which governs the long-distance behavior of the local
order-parameter correlations $\langle Q_i Q_j \rangle$. The location
of the maximum in $X^Q_L$ also shifts with $L$, which gives further important 
information about the critical exponents.

\begin{figure}
\center
\epsfxsize=5.5cm \epsfysize=5.5cm \epsfbox{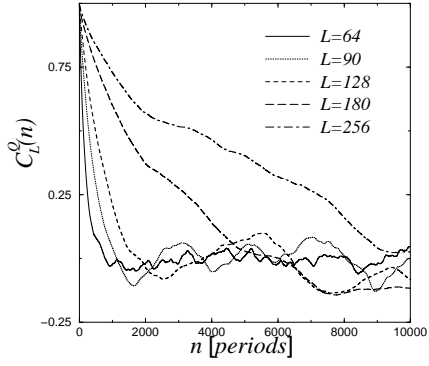}
\caption{Critical slowing down for the order parameter at 
$T$$=$$0.8T_{\rm c}$ and $H_0$$=$$0.3J$ at $\Theta$$=$$\Theta_{\rm c}$,
as shown by the normalized autocorrelation function $C^{Q}_{L}(n)$.}
\label{crit_slow}
\end{figure}

The normalized stationary time-displaced autocorrelation function of the order
parameter,
\begin{equation}
C^{Q}_{L}(n) = \frac{\langle Q(j) Q(j+n)\rangle -  \langle Q(j) \rangle^{2}}
                   {\langle Q^{2}(j)\rangle -  \langle Q(j) \rangle^{2}} \;,
\label{auto_corr}
\end{equation} 
provides further insights into the DPT as the system exhibits critical slowing
down [Fig.~\ref{crit_slow}]. This can be seen as increasing correlation times
with increasing system sizes. In Sec.~III.D we provide a quantitative
analysis of the correlation times.

\begin{figure}
\center
\epsfxsize=6cm \epsfysize=6cm \epsfbox{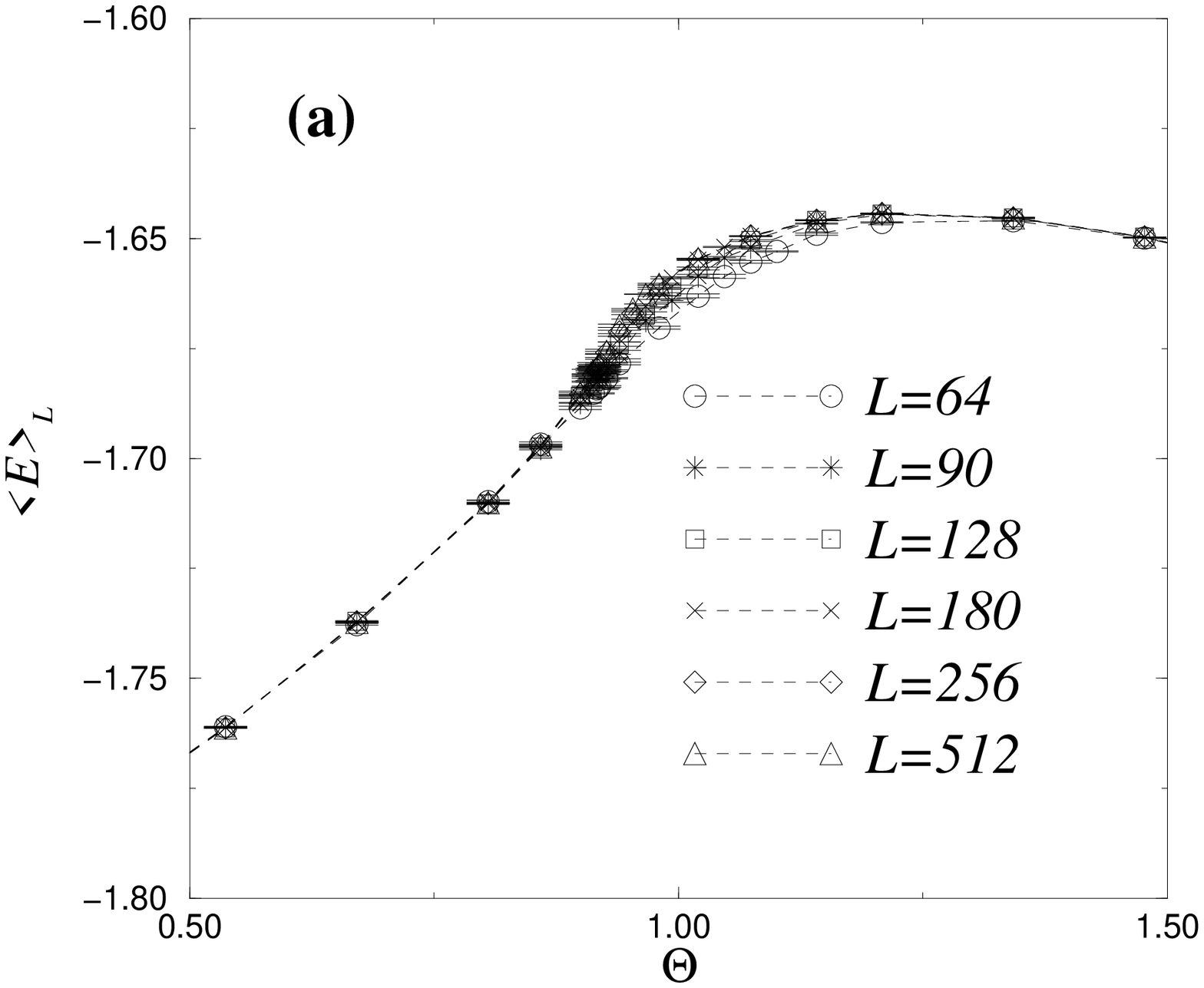}
\epsfxsize=6cm \epsfysize=6cm \epsfbox{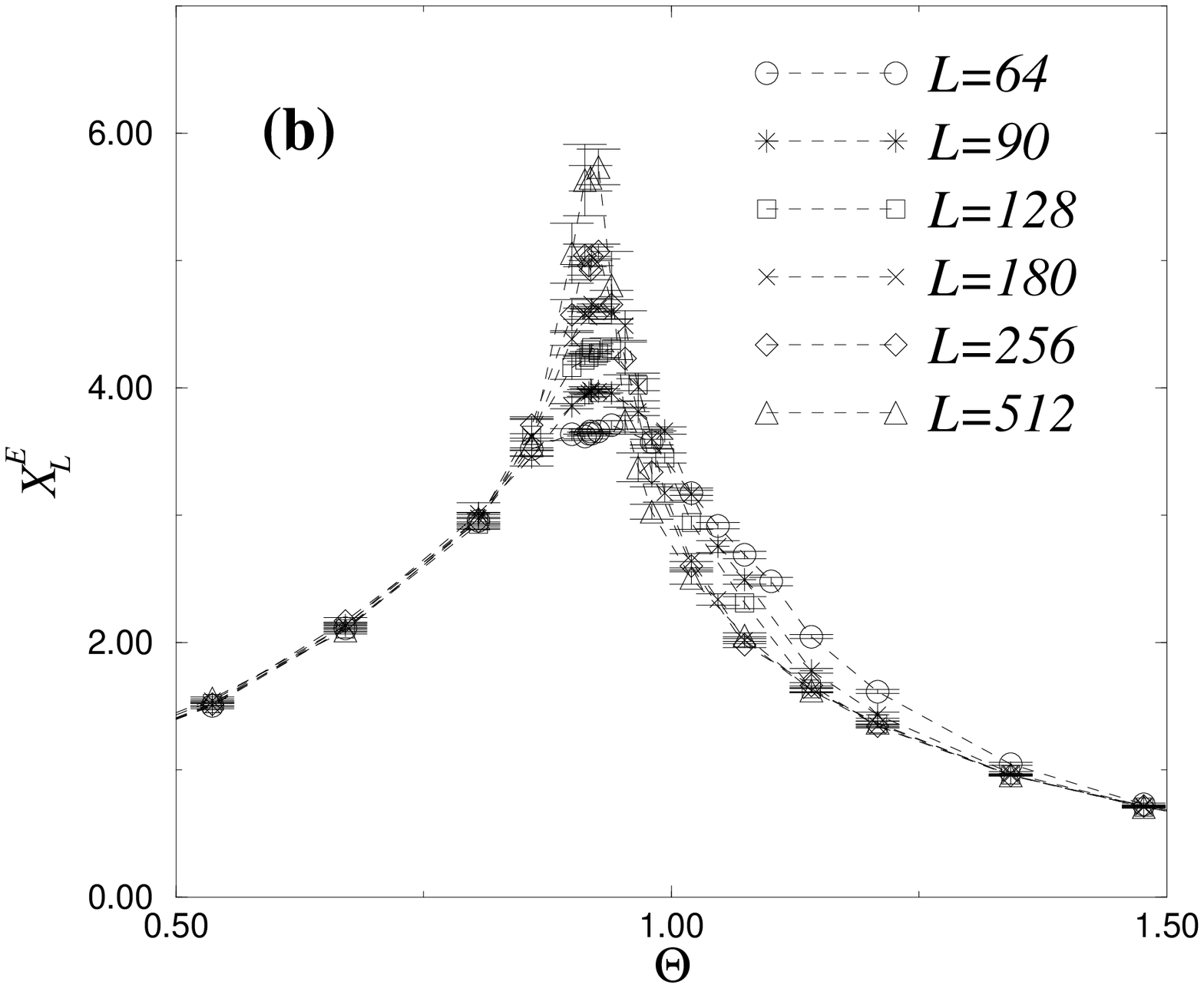}
\caption{Finite-size behavior of the period-averaged internal energy 
at $T$$=$$0.8T_{\rm c}$ and $H_0$$=$$0.3J$ for various system sizes.
(a) The period-averaged internal energy.
(b) The scaled energy variance, $X^E_L$ as defined in 
Eq. (\protect\ref{eq:XE}).}
\label{E_raw}
\end{figure}

We also measured the period-averaged internal energy [Eq.~(\ref{energy_def})] 
and its fluctuations \cite{ACHA97D}
\begin{equation}
X^{E}_{L}=L^2 \left(\langle E^2 \rangle_L - \langle E \rangle_L ^2 \right) \;,
\label{eq:XE}
\end{equation}
as can be seen in Fig.~\ref{E_raw}.
The peaks of these fluctuations exhibit a slow increase with the system size 
(compared to the order-parameter fluctuations), as one may anticipate by
analogy with the equilibrium heat capacity.


\subsection{Finite-size scaling}

\begin{figure}
\center
\epsfxsize=6cm \epsfysize=6cm \epsfbox{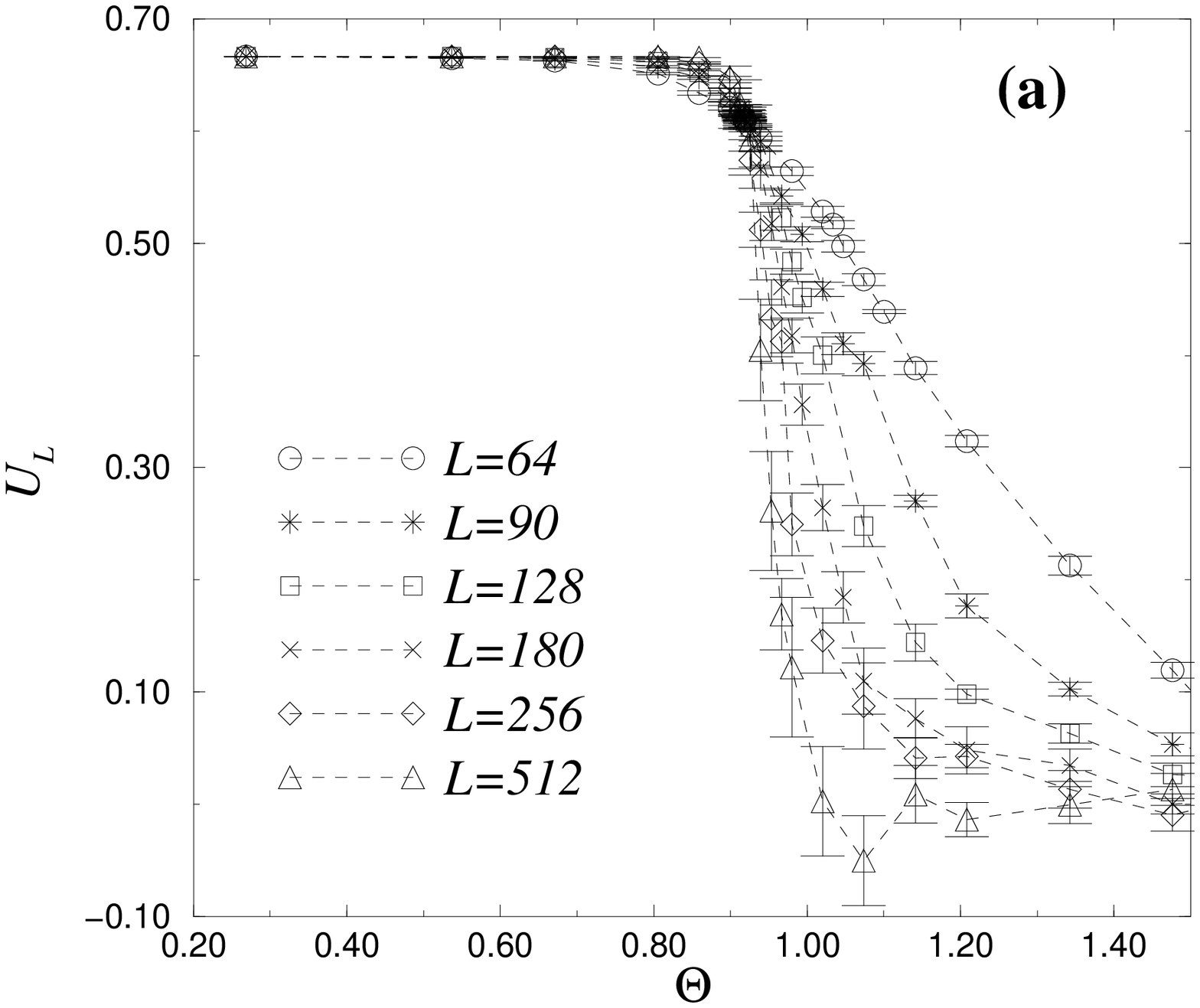}
\epsfxsize=6cm \epsfysize=6cm \epsfbox{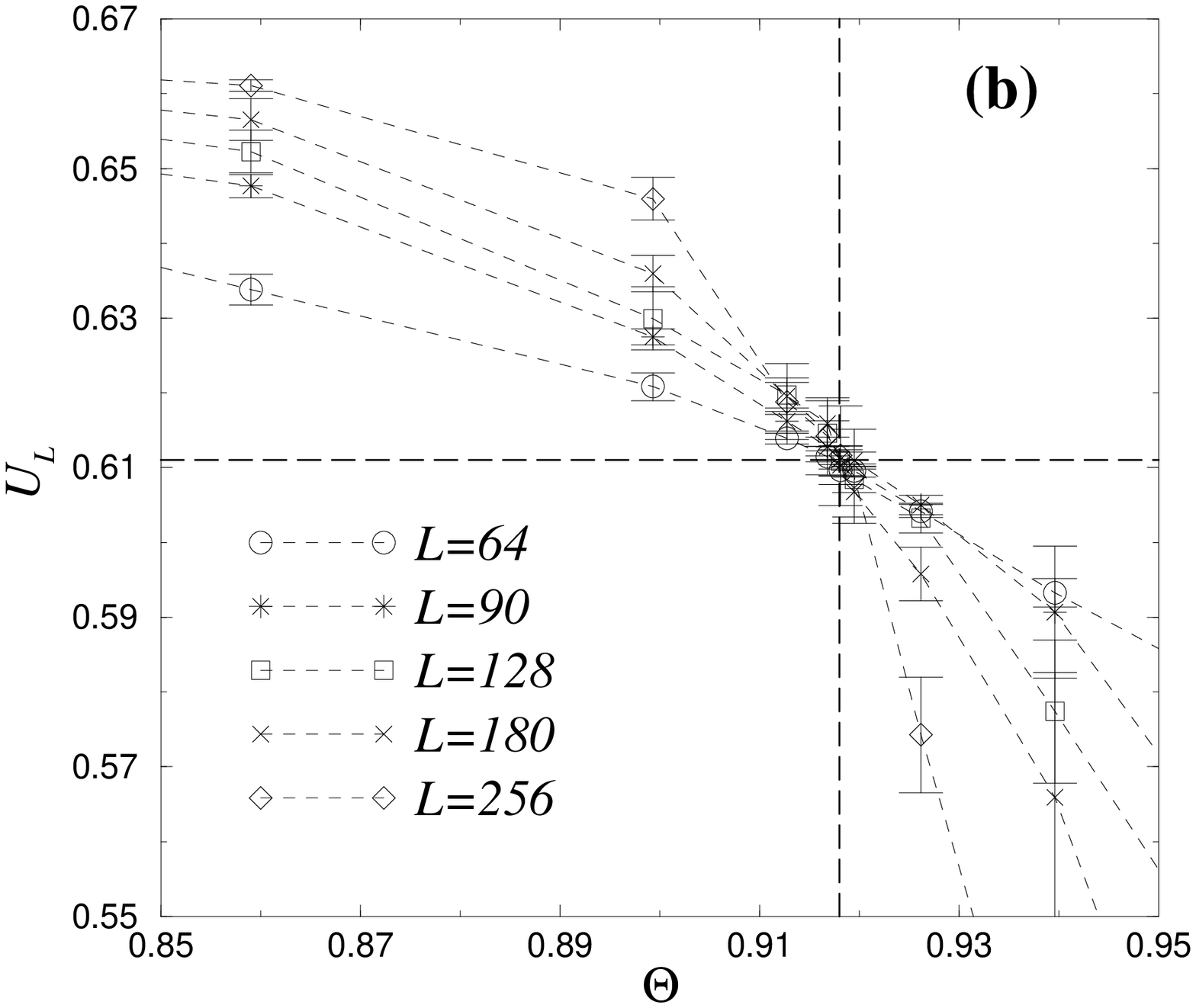}
\caption{(a) The fourth-order cumulant as defined in 
Eq.~(\protect\ref{eq:cumulant}) 
at $T$$=$$0.8T_{\rm c}$ and $H_0$$=$$0.3J$ for various system sizes.
(b) The region around the cumulant crossing in (a) enlarged.
The horizontal and vertical dashed lines indicate the fixed-point value
$U^{*}$$=$$0.611$ and the scaled critical half-period 
$\Theta_{\rm c}$$=$$0.918$, respectively.}
\label{fig_cumul}
\end{figure}

Scaling laws and finite-size scaling for equilibrium systems
with an a-priori known Hamiltonian can be systematically derived using the 
concepts of the free energy and the renormalization group \cite{GOLD}. 
The kinetic Ising model with the explicitly time-dependent Hamiltonian,
Eq.~(\ref{eq:Hamil}), is driven far from equilibrium. 
Although the order-parameter distribution $P(Q)$ is stationary, the effective 
Hamiltonian controlling its fixed-point behavior has not been known until
recently \cite{Fuji} (after the completion of this study).
Motivated by the similarity of the finite-size effects shown in 
Figs.~\ref{Q_raw}-\ref{E_raw} to those characteristic of a 
typical continuous phase transition, we borrow the corresponding 
scaling assumptions from equilibrium finite-size scaling.
For our model the quantity analogous to the reduced temperature in 
equilibrium systems (i.e., the distance from the {\em infinite}-system 
critical point) is
\begin{equation}
\theta= \frac{|\Theta-\Theta_{\rm c}|}{\Theta_{\rm c}}\;.
\label{eq:reduced_Theta}
\end{equation}
Finite-size scaling theory provides simple scaling relations for the
observables for finite systems in the critical regime \cite{BIND92,BIND90}:
\begin{eqnarray}
\langle |Q|\rangle _L & = & 
L^{-\beta /\nu} {\cal F}_{\pm}(\theta L^{1/\nu}) \label{full_scaling_Q}\\ 
X^Q_L & = & 
L^{\gamma /\nu} {\cal G}_{\pm}(\theta L^{1/\nu}) \label{full_scaling_XQ} \\
X^E_L & = & c_1\ln\left( L{\cal J}_{\pm}(\theta L^{1/\nu})\right) \;,
\label{full_scaling_XE}
\end{eqnarray}
where ${\cal F}_{\pm}$, ${\cal G}_{\pm}$, and ${\cal J}_{\pm}$ are 
scaling functions and the $+$ ($-$) 
index refers to \mbox{$\Theta >\Theta_{\rm c}$} 
\mbox{($\Theta <\Theta_{\rm c}$)}.
The logarithmic scaling in $X^E_L$ is motivated by the very slow divergence
of the scaled period-averaged energy variance [Fig.~\ref{E_raw}(b)]. 
The above formulation of scaling is explicitly based on the infinite-system 
critical point $\Theta_{\rm c}$, which can be estimated with far greater 
accuracy than the location of the maximum of the order-parameter fluctuations 
for the individual finite system sizes.  
We use the fourth-order cumulant intersection method \cite{BIND92,BIND90} 
to estimate the value  of $\Theta_c$ at which the transition occurs in an 
{\em infinite} system. In order to do this, we plot  
\begin{equation} 
U_L=1 - \frac{\langle Q^4\rangle_L}{3\langle Q^2\rangle_L^2} \;
\label{eq:cumulant}
\end{equation}
as a function of $\Theta$ for several system sizes as shown in 
Fig.~\ref{fig_cumul}. For the largest system ($L$=$512$) 
the statistical uncertainty in $U_L$ was too large to 
use it to obtain estimates for the crossing. Our estimate for the 
dimensionless critical half-period, based on the remaining five system sizes,
is $\Theta_{\rm c}$$=$$0.918$$\pm$$0.005$ with a fixed-point value 
$U^*$$=$$0.611$$\pm$$0.003$ for the cumulant [Fig~\ref{fig_cumul}(b)].

\begin{figure}
\center
\epsfxsize=5.5cm \epsfysize=5.5cm \epsfbox{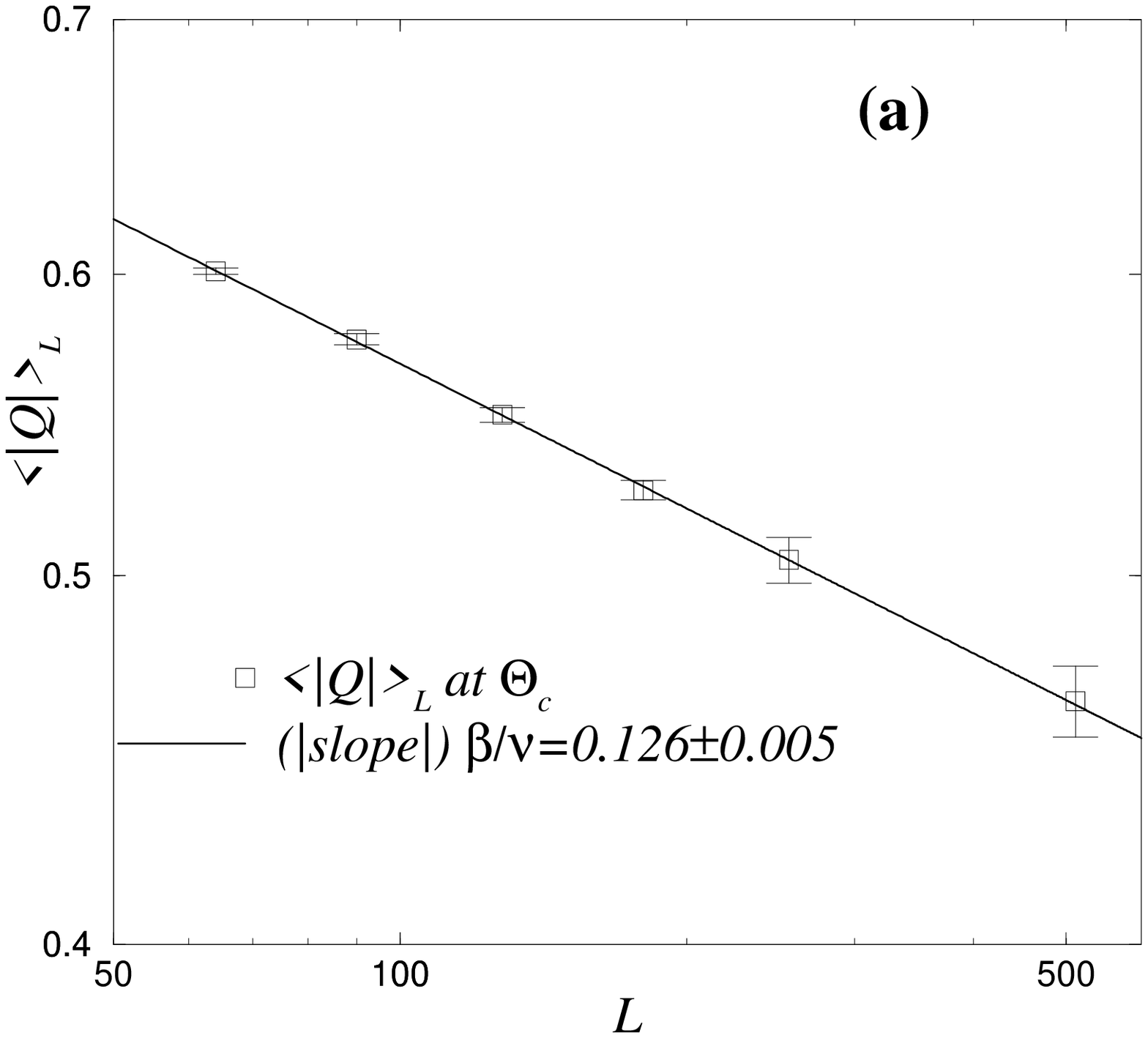}
\epsfxsize=5.5cm \epsfysize=5.5cm \epsfbox{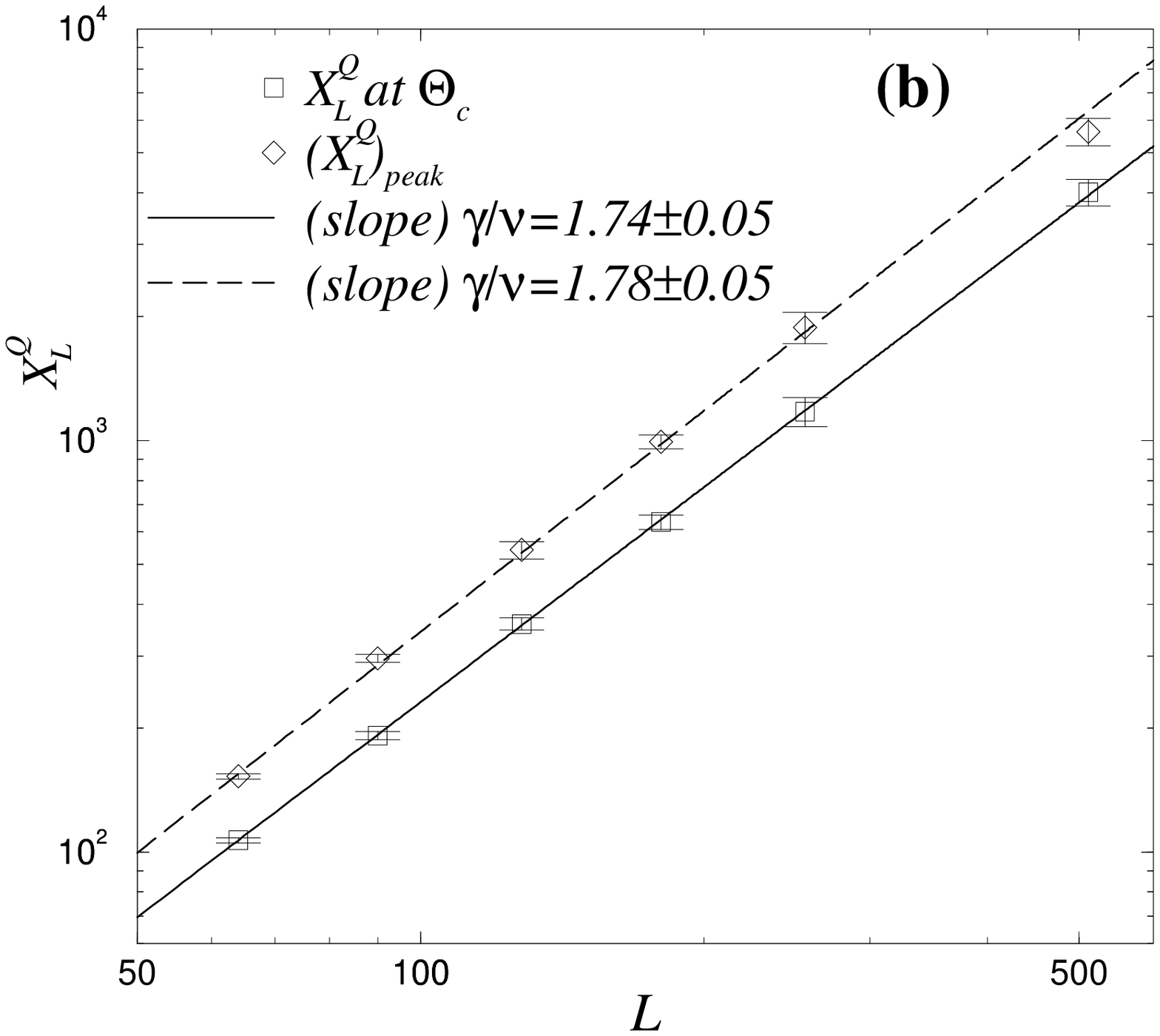}
\epsfxsize=5.5cm \epsfysize=5.5cm \epsfbox{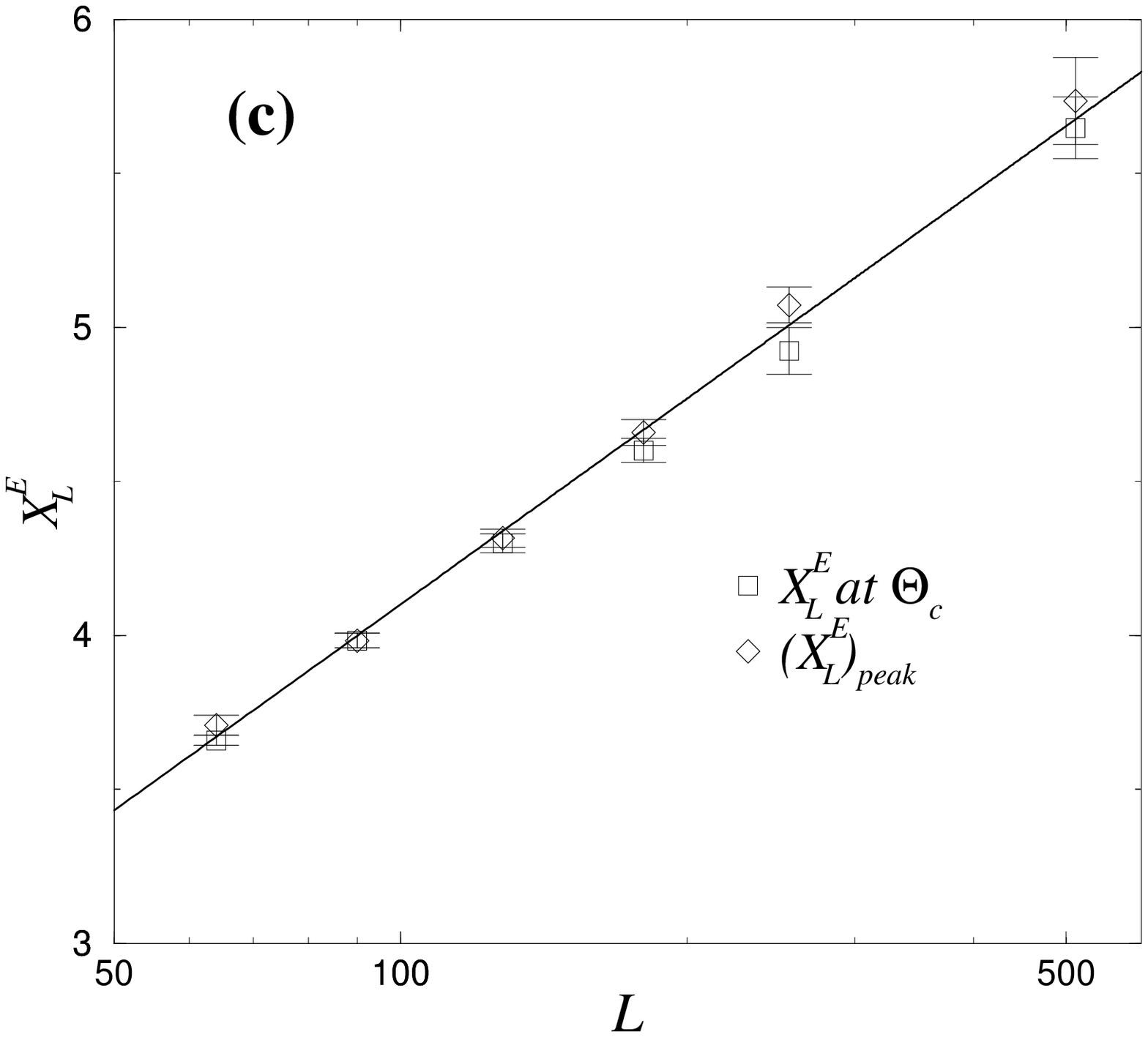}
\caption{Critical exponent estimates at $T$$=$$0.8T_{\rm c}$ and 
$H_0$$=$$0.3J$. Straight lines are the weighted least-square fits.
(a) Determining $\beta/\nu$ through the finite-size effects of the order 
parameter, based on Eq.~(\ref{scaling_Q}) (log-log plot). 
(b) Determining $\gamma/\nu$ through the finite-size effects of the 
order-parameter fluctuations, based on Eq.~(\ref{scaling_XQ}) (log-log plot).
(c) Showing the  logarithmic divergence of the period-averaged energy 
fluctuations, based on Eq.~(\ref{scaling_XE}) (log-lin plot).}
\label{fit}
\end{figure}

Then at $\Theta_c$ the scaling forms 
Eqs.~(\ref{full_scaling_Q}-\ref{full_scaling_XE}) yield
\begin{eqnarray}
\langle |Q|\rangle _L & \propto & 
L^{-\beta /\nu} \label{scaling_Q}\\ 
X^Q_L & \propto & 
L^{\gamma /\nu} \label{scaling_XQ} \\
X^E_L & \propto & c_2 + c_1\ln(L) \;,
\label{scaling_XE}
\end{eqnarray}
which enable us to estimate the exponent ratios $\beta/\nu$ and $\gamma/\nu$,
and to directly check the postulated logarithmic divergence in the 
period-averaged energy fluctuations.
Plotting $\langle |Q|\rangle _L$ and $X^Q_L$ at $\Theta_c$ and utilizing
a weighted linear least-squares fit to the logarithmic data yields
$\beta/\nu$$=$$0.126\pm 0.005$ [Fig.~\ref{fit}(a)] and 
$\gamma/\nu$$=$$1.74\pm 0.05$ [Fig.~\ref{fit}(b)]. Note that these values are 
extremely close (within statistical errors) to the corresponding ratios for
the {\em equilibrium} two-dimensional Ising universality class, 
$\beta/\nu$$=$$1/8$$=$$0.125$ and $\gamma/\nu$$=$$7/4$$=$$1.75$.
Further, the straight line in Fig.~\ref{fit}(c) indicates the slow
logarithmic divergence of $X^E_L$ at the critical point. 
In addition to the scaling at $\Theta_c$, we also checked the divergences
of the peaks of the fluctuations, $(X^Q_L)_{\rm peak}$ and 
$(X^E_L)_{\rm peak}$, 
since they asymptotically should follow the same scaling laws, 
Eqs.~(\ref{scaling_XQ}) and (\ref{scaling_XE}), respectively. The measured 
exponent $\gamma/\nu$$=$$1.78\pm 0.05$ for $(X^Q_L)_{\rm peak}$ 
and the logarithmic divergence for $(X^E_L)_{\rm peak}$ agree to within the 
statistical errors with the results obtained at $\Theta_c$, as can be seen in
Fig.~\ref{fit}(b) and (c), respectively.  

\begin{figure}
\center
\epsfxsize=6cm \epsfysize=6cm \epsfbox{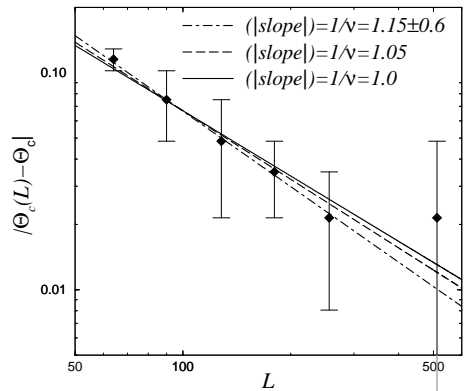}
\caption{Exponent estimate for $\nu$ at $T$$=$$0.8T_{\rm c}$ and 
$H_0$$=$$0.3J$, based on Eq.~(\ref{scaling_Tc}) (log-log plot). The dot-dashed
line is a weighted least-squares fit (excluding the smallest and the largest
system) yielding the slope 
$1/\nu$$=$$1.15\pm0.6$ ($\nu$$=$$0.87\pm0.4$). 
The dashed line represents the ``optimal'' value for 
this exponent, using the best quality data collapse for the scaling function
Eq.~(\protect\ref{full_scaling_Q}) as discussed in the text, yielding 
$1/\nu$$=$$1.05$ ($\nu$$=$$0.95$). The solid line represents the 
two-dimensional equilibrium Ising exponent $\nu$$=$$1.0$.}
\label{Tc_scale}
\end{figure}

From the finite-system shifting of the transition one can estimate the 
correlation-length exponent $\nu$ by tracking the shift in the location 
of the maximum in $X^Q_L$:
\begin{equation}
|\Theta_{c}(L)-\Theta_{c}| \propto L^{-1/\nu}\;,
\label{scaling_Tc}
\end{equation}
where $\Theta_{c}(L)$ is the location of the peak for finite systems.
However, the precision of this method for our data is very poor, due to 
limited resolution in finding the locations of the maxima and consequently the 
large relative errors in $|\Theta_{c}(L)-\Theta_{c}|$. Excluding the smallest 
(due to strong corrections to scaling) and the largest systems (due to very 
poor resolution and extremely large statistical error), we obtain 
$\nu$$=$$0.87\pm0.4$, but the large error estimate obviously implies rather 
poor accuracy [Fig.~\ref{Tc_scale}].

\begin{figure}
\center
\epsfxsize=6cm \epsfysize=6cm \epsfbox{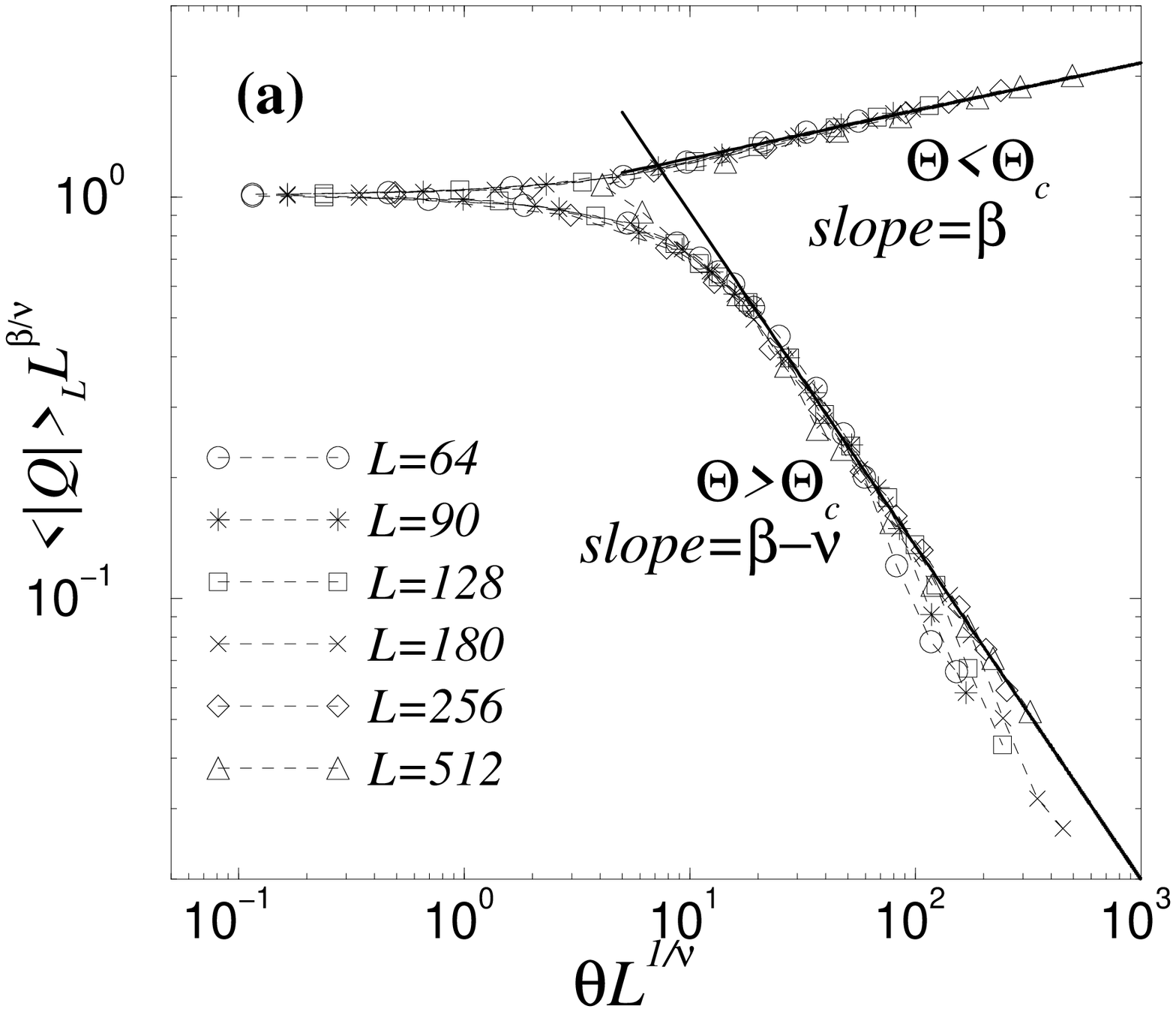}
\epsfxsize=6cm \epsfysize=6cm \epsfbox{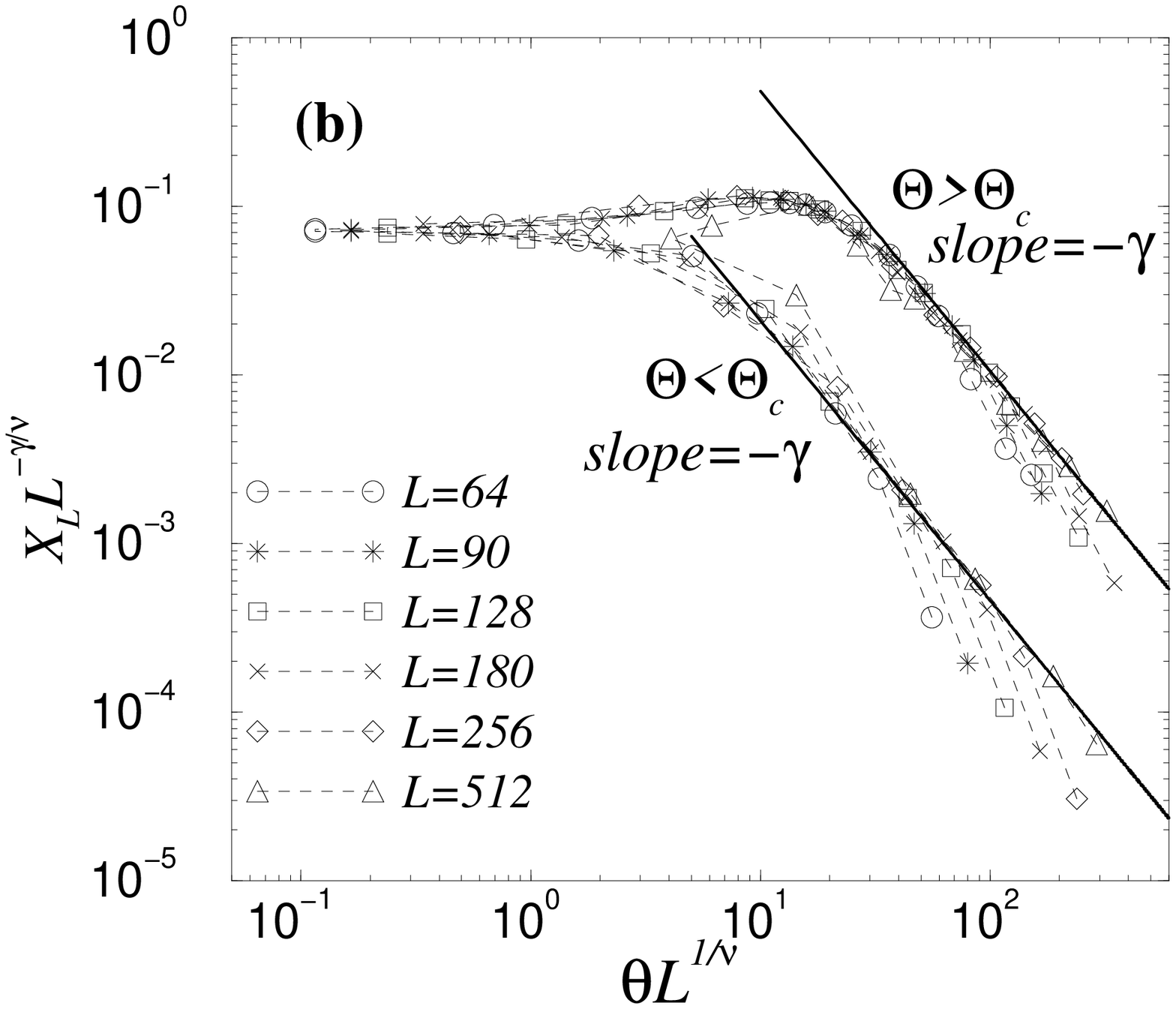}

\epsfxsize=6cm \epsfysize=6cm \epsfbox{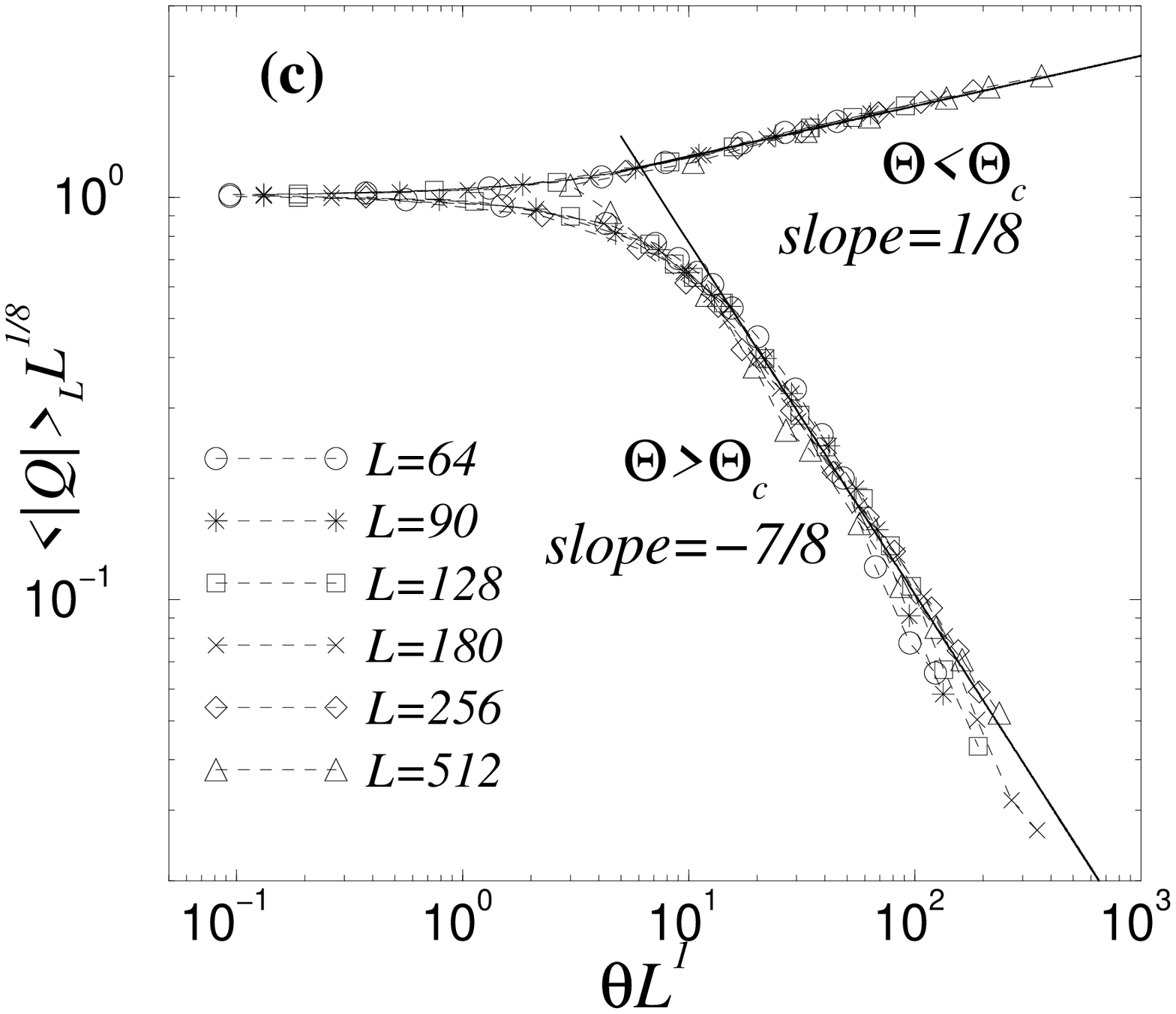}
\epsfxsize=6cm \epsfysize=6cm \epsfbox{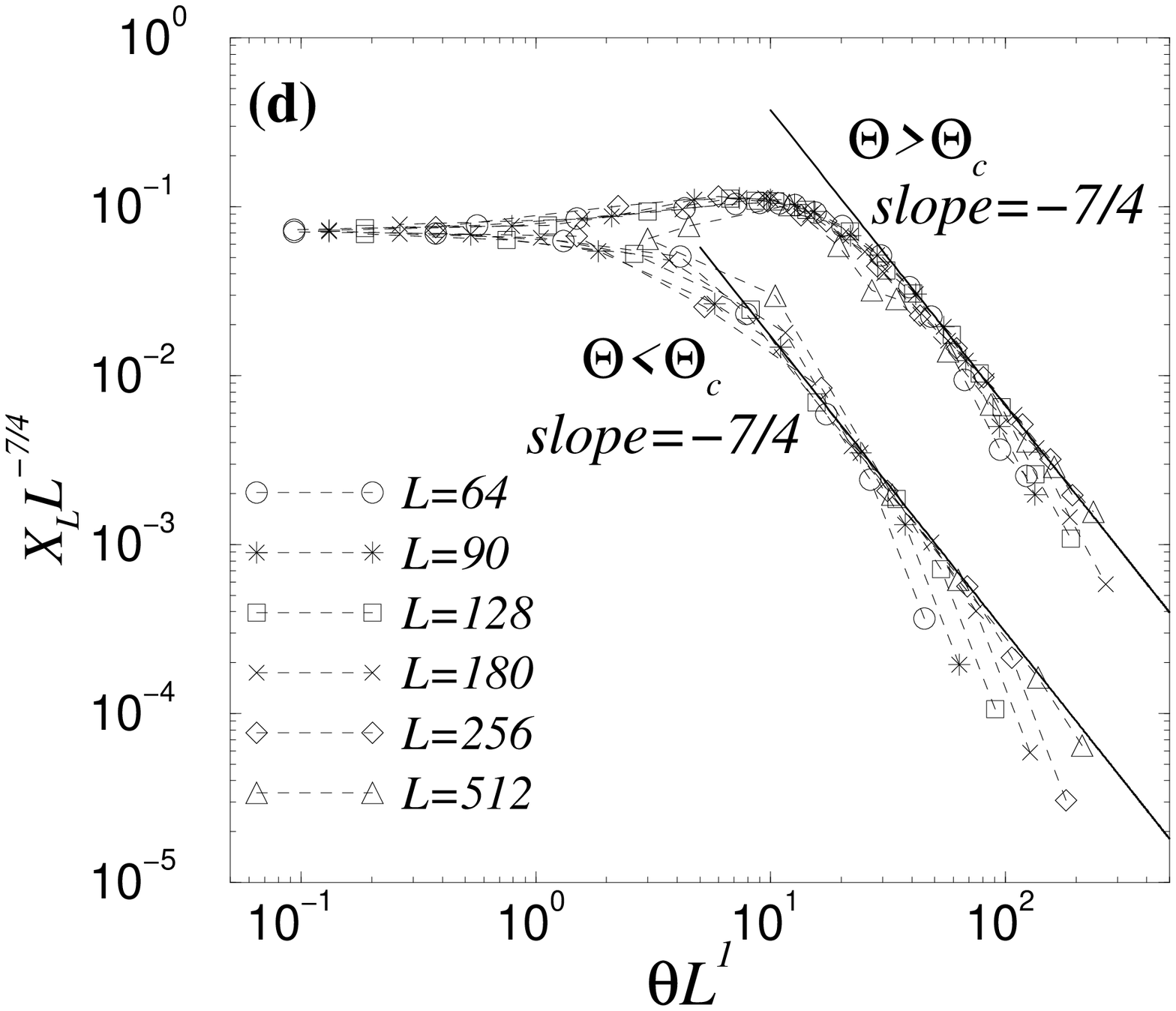}
\caption{Finite-size scaling (full data collapse) at $T$$=$$0.8T_{\rm c}$ and
$H_0$$=$$0.3J$ using $\beta/\nu$$=$$1/8$, $\gamma/\nu$$=$$7/4$ 
(two-dimensional equilibrium Ising values), and $\nu$$=$$0.95$ 
(which yields the best quality data collapse).
(a) For the order parameter $\langle|Q|\rangle_L$ (log-log plot).
(b) For the scaled order-parameter variance $X^Q_L$ (log-log plot).
Straight lines in both graphs represent the asymptotic large-argument 
behaviors of the scaling functions ${\cal F}_{\pm}$ and ${\cal G}_{\pm}$ given 
in Eqs.~(\ref{full_scaling_Q}) and (\ref{full_scaling_XQ}), respectively.
Figures (c) and (d) are the same as (a) and (b), except that the exact Ising
exponent $\nu$$=$$1.0 $ is used. }
\label{full_collapse}
\end{figure}

To obtain a more complete picture of how well the scaling relations in
Eqs.~(\ref{full_scaling_Q}) and (\ref{full_scaling_XQ}) hold, we plot 
$\langle |Q|\rangle _L L^{\beta/\nu}$ [Fig.~\ref{full_collapse}(a)] and 
$X^Q_LL^{-\gamma/\nu}$ [Fig.~\ref{full_collapse}(b)] vs
$\theta L^{1/\nu}$ \cite{LANDAU76}. For the exponent ratios we used
$\beta/\nu$$=$$1/8$$=$$0.125$, and $\gamma/\nu$$=$$7/4$$=$$1.75$, since our 
estimate for those (within small statistical errors) implied that they take 
on the equilibrium two-dimensional Ising universal values.
Most importantly, we used various values of $\nu$ between $0.5$ and $1.2$ to 
find the best data collapse as observed visually, since our estimate for this 
exponent was far from reliable. The ``optimal'' value obtained this way
(by showing scaling plots to group members who did not know 
the particular values of $\nu$ used), and used in 
Fig.~\ref{full_collapse}(a) and (b), is $\nu$$=$$0.95\pm0.15$. 
Full scaling plots using the exact Ising exponents are also
shown in Fig.~\ref{full_collapse}(c) and (d), and they result in similarly
good data collapse. 

\subsection{Order-parameter histograms at criticality}

\begin{figure}
\center
\epsfxsize=6cm \epsfysize=6cm \epsfbox{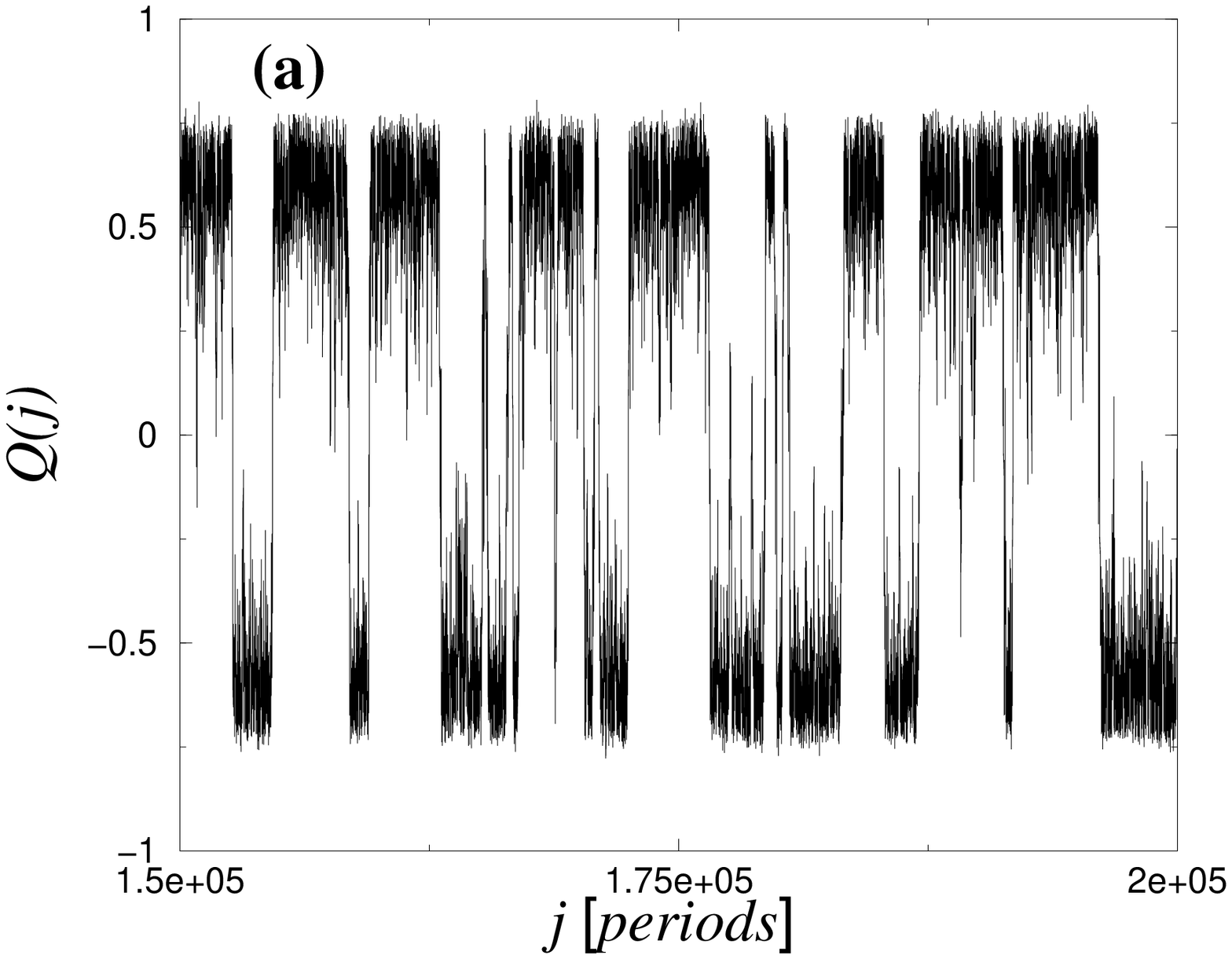}
\epsfxsize=6cm \epsfysize=6cm \epsfbox{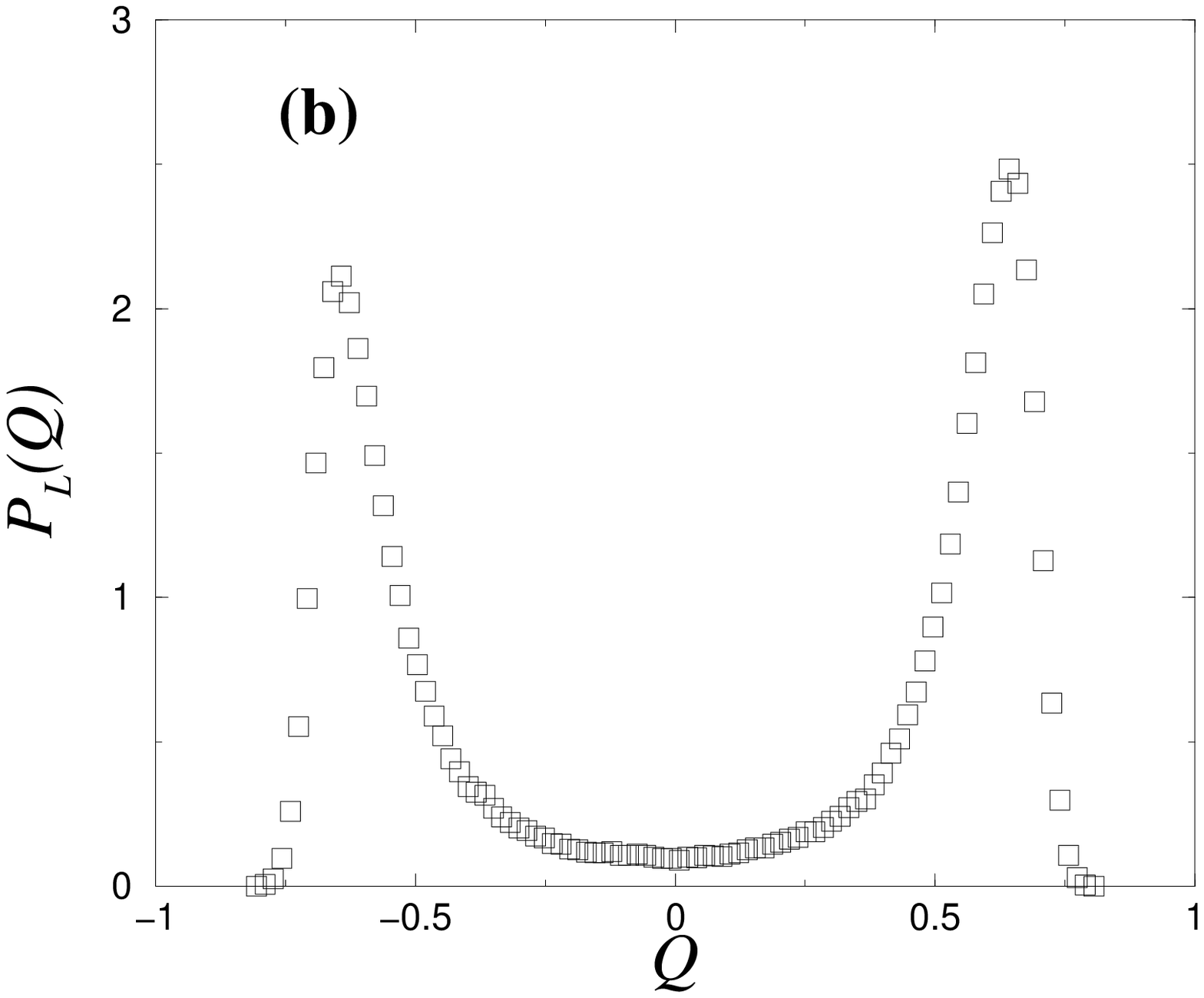}
\caption{(a) Short segment of the order-parameter time series at 
$T$$=$$0.8T_{\rm c}$ and $H_0$$=$$0.3J$ for an $L$$=$$128$ system at 
$\Theta_{\rm c}$ \protect\cite{endnote1}. 
(b) Order-parameter histogram for the same parameters.}
\label{series_hist}
\end{figure}

\begin{figure}
\center
\epsfxsize=5.5cm \epsfysize=5.5cm \epsfbox{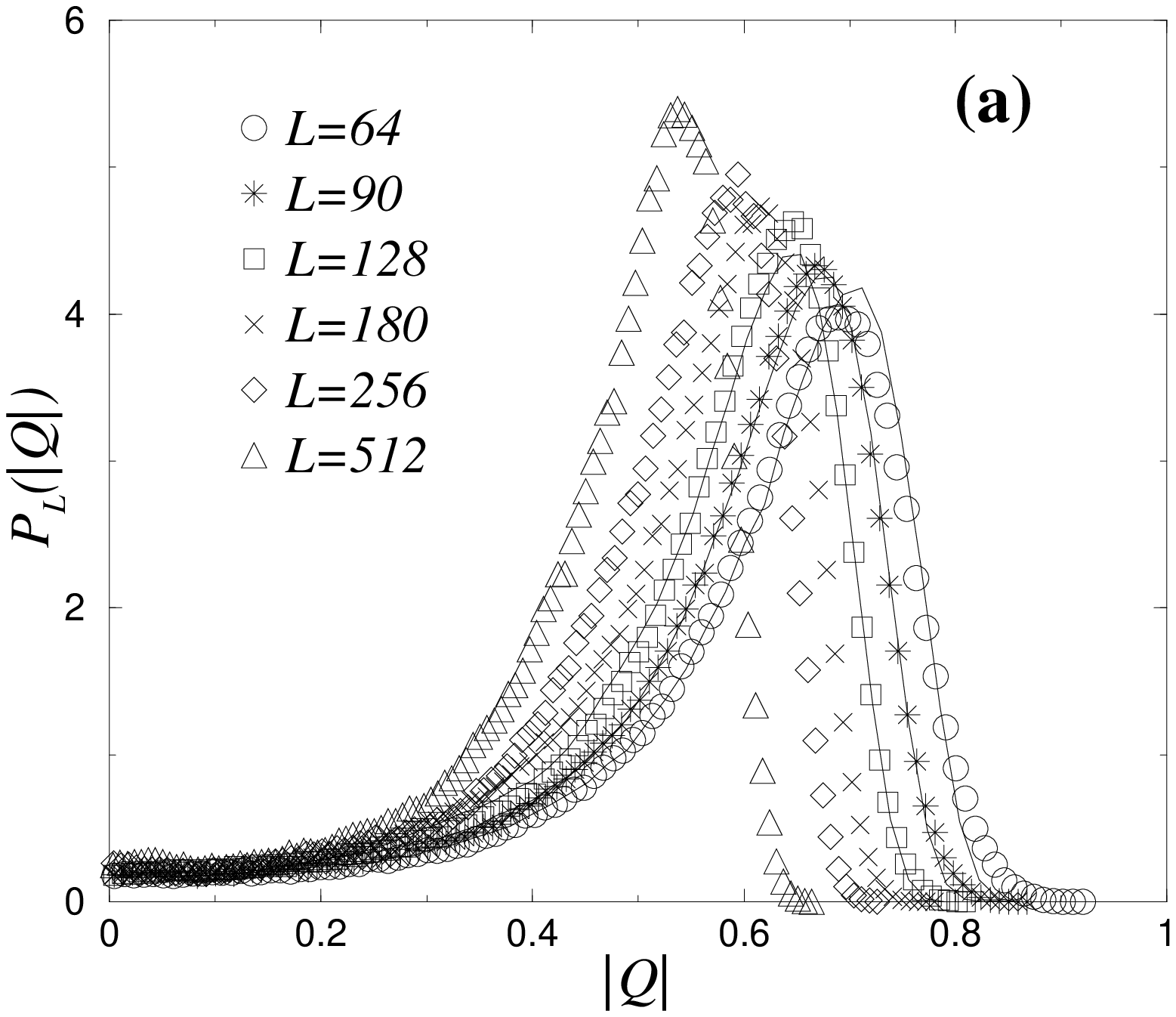}
\epsfxsize=5.5cm \epsfysize=5.5cm \epsfbox{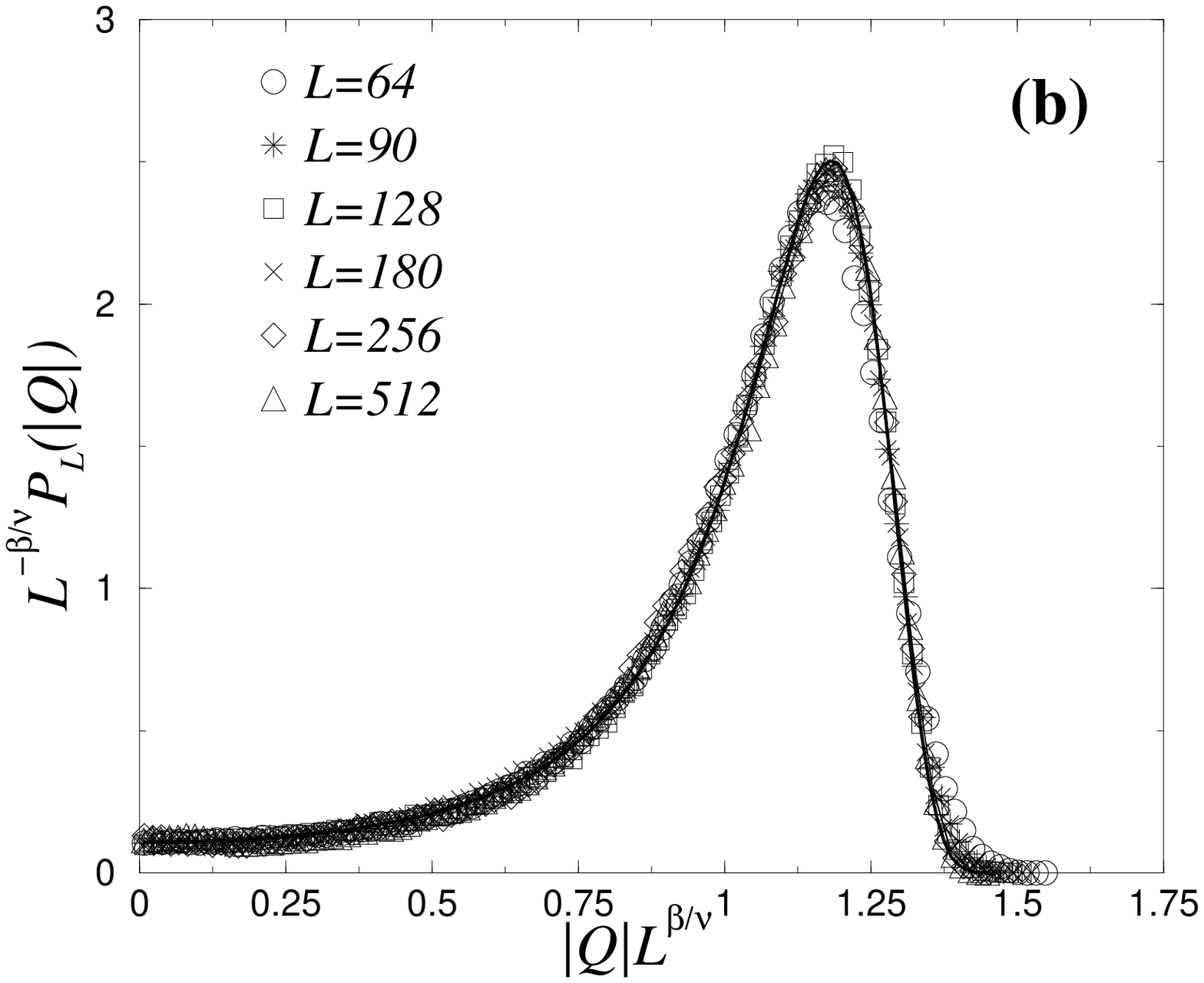}
\epsfxsize=5.5cm \epsfysize=5.5cm \epsfbox{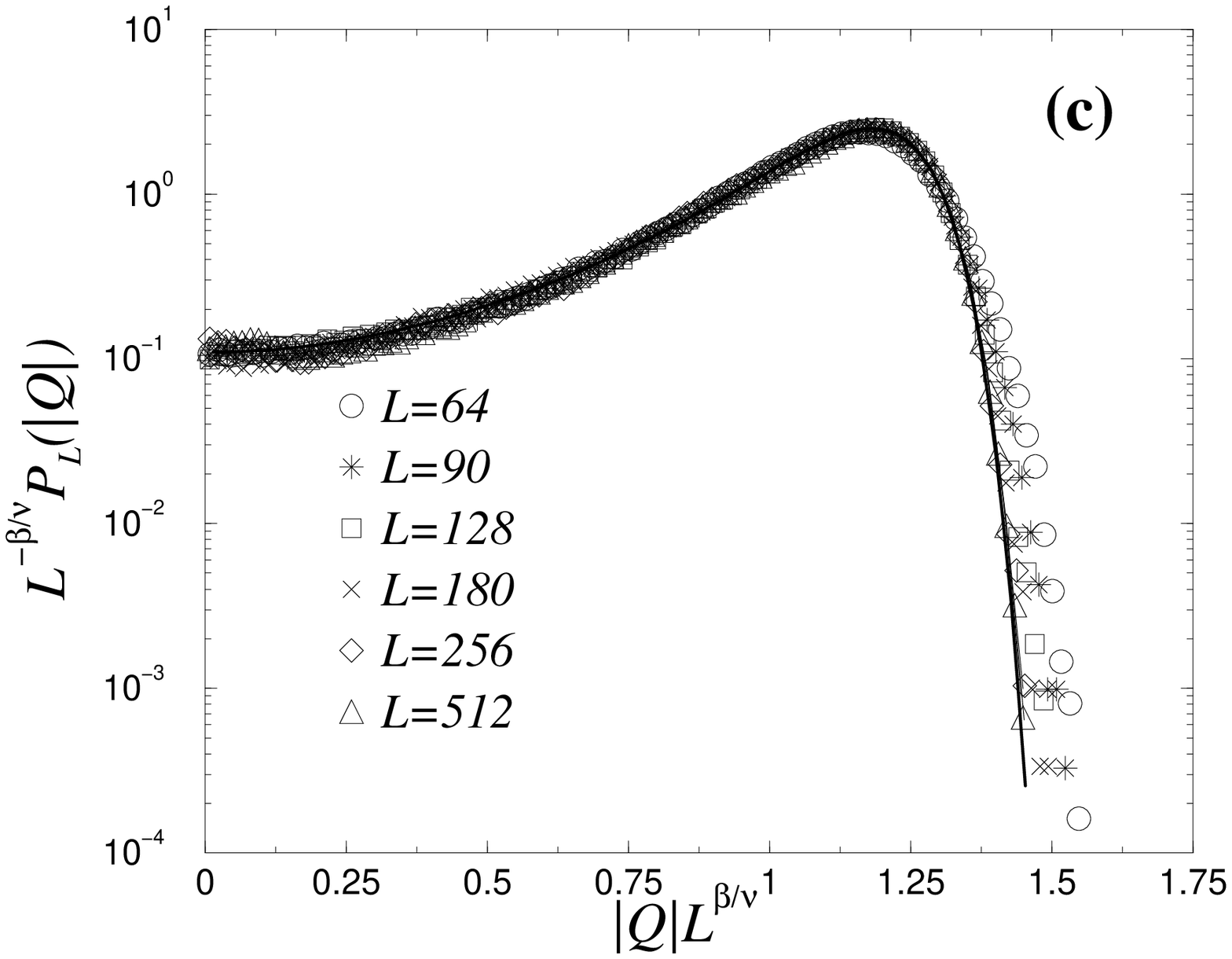}
\caption{Order-parameter histograms $P_{L}(|Q|)$ at $T$$=$$0.8T_{\rm c}$ and 
$H_0$$=$$0.3J$ for various system sizes at $\Theta_{\rm c}$.
The thin solid lines represent two-dimensional {\em equilibrium} Ising 
order-parameter histograms at the critical point for $L$$=$$64$, $90$, and 
$128$ on all three graphs. 
(a) Order-parameter distributions. 
(b) Scaled order-parameter distributions, according to Eq.~(\ref{scaling_PQ})
with $\beta/\nu$$=$$1/8$. 
The bold solid line is the corresponding (Monte Carlo) two-dimensional 
equilibrium Ising distribution without 
any additional scaling parameters \protect\cite{Janke}.
(c) Same as (b) on lin-log scales to enhance the view of the 
corrections to scaling for small systems at large $|Q|$.}
\label{scaled_hist}
\end{figure}

We devote this subsection to analyzing the universal characteristics of the
full order-parameter distribution, $P(Q)$, at the critical point. This 
distribution is bimodal for finite systems if observed for sufficiently long 
times (Fig.~\ref{series_hist}) \cite{endnote1}.
It is more convenient to focus on the distribution of $|Q|$, avoiding the 
effect of the insufficient number of switching events between the two 
symmetry-broken phases for large systems, which causes the skewness in 
Fig.~\ref{series_hist}(b).
Figure~\ref{scaled_hist}(a) shows the 
order-parameter distributions $P_{L}(|Q|)$ at the critical point for various 
system sizes. Finite-size scaling arguments \cite{BIND92,BIND90} suggest that 
at $\Theta_{\rm c}$
\begin{equation}
P_{L}(|Q|)=L^{\beta/\nu}{\cal P}(L^{\beta/\nu} |Q|) \;.
\label{scaling_PQ}
\end{equation}
Thus, the scaled distributions, $L^{-\beta/\nu}P_{L}(|Q|)$ vs 
$x$$=$$|Q|L^{\beta/\nu}$, should fall on the same curve ${\cal P}(x)$ for 
different system sizes. 
Again, we used $\beta/\nu=1/8$. The quality of the data collapse is quite 
impressive [Fig.~\ref{scaled_hist}(b)], with deviations mainly observed for 
the smallest $L$ and the largest values of $|Q|$ [Fig.~\ref{scaled_hist}(c)], 
possibly as a result of corrections to scaling.

What we find somewhat surprising, is that the distribution appears to be 
identical (except for stronger corrections to scaling at the DPT) to that of 
the equilibrium two-dimensional Ising model on a square lattice with periodic 
boundary conditions at criticality, 
{\em without} a need for any additional scaling parameters. 
We checked this by performing standard equilibrium two-dimensional Ising 
simulations with Glauber dynamics and system sizes ranging from $L$$=$$64$ to 
$L$$=$$128$, and also by comparing our scaled DPT order-parameter histograms 
to the high-precision two-dimensional equilibrium Ising MC data of Ref. 
\cite{Janke} (Fig.~\ref{scaled_hist}). We had expected the 
{\em shapes} of the distributions to be identical for the DPT and equilibrium 
Ising model, as a consequence of the identical values for the cumulant 
fixed-point value $U^*$. However, it is not obvious 
to us why the microscopic length scales in the DPT and the equilibrium Ising 
model also appear to be identical, as evidenced by the absence of the need for
an additional scaling, $L$$\rightarrow$$L/a$ ($a$ being the microscopic 
length scale in the DPT).

\subsection{Critical slowing down}

\begin{figure}
\center
\epsfxsize=5.5cm \epsfysize=5.5cm \epsfbox{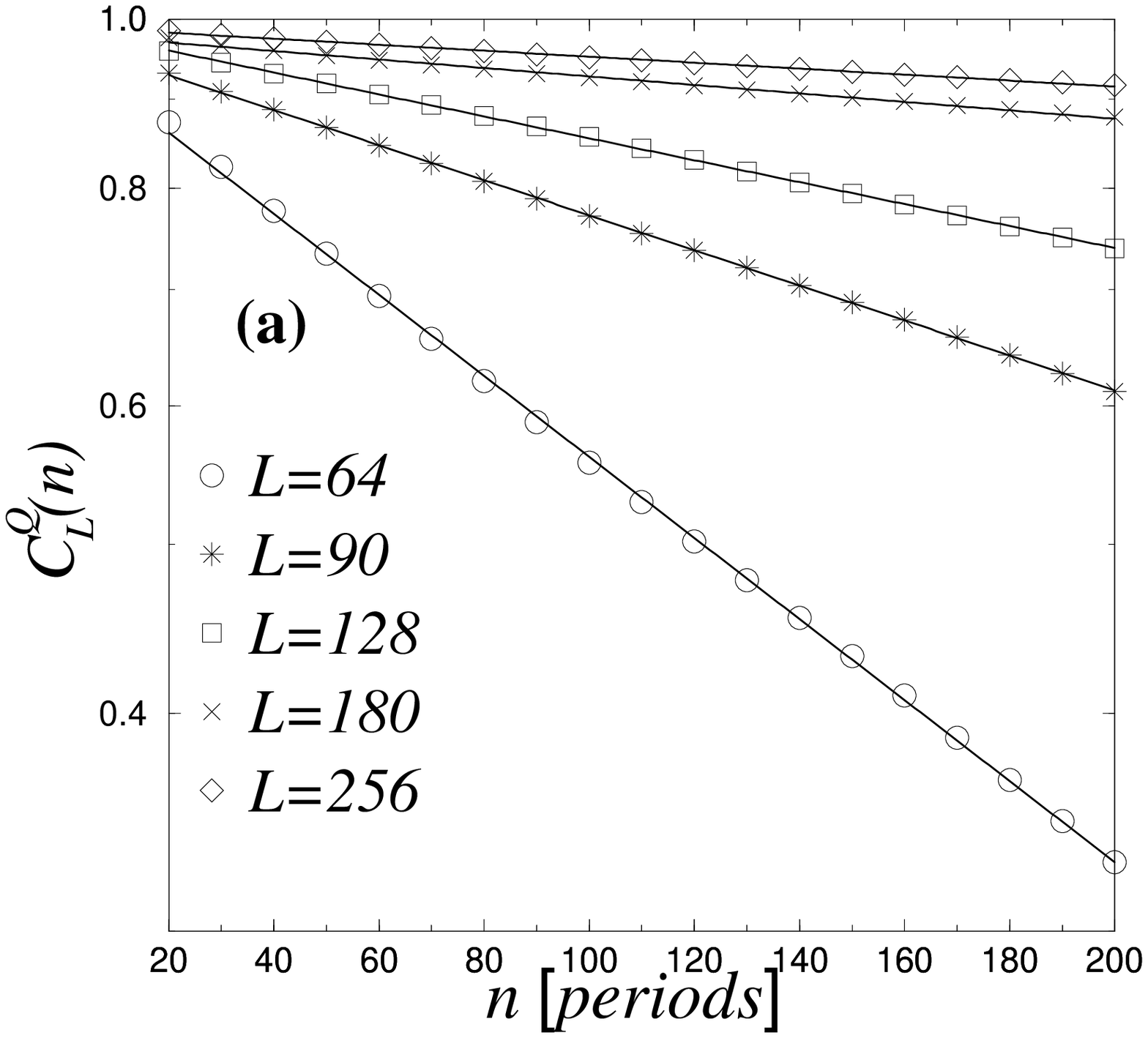}
\epsfxsize=5.5cm \epsfysize=5.5cm \epsfbox{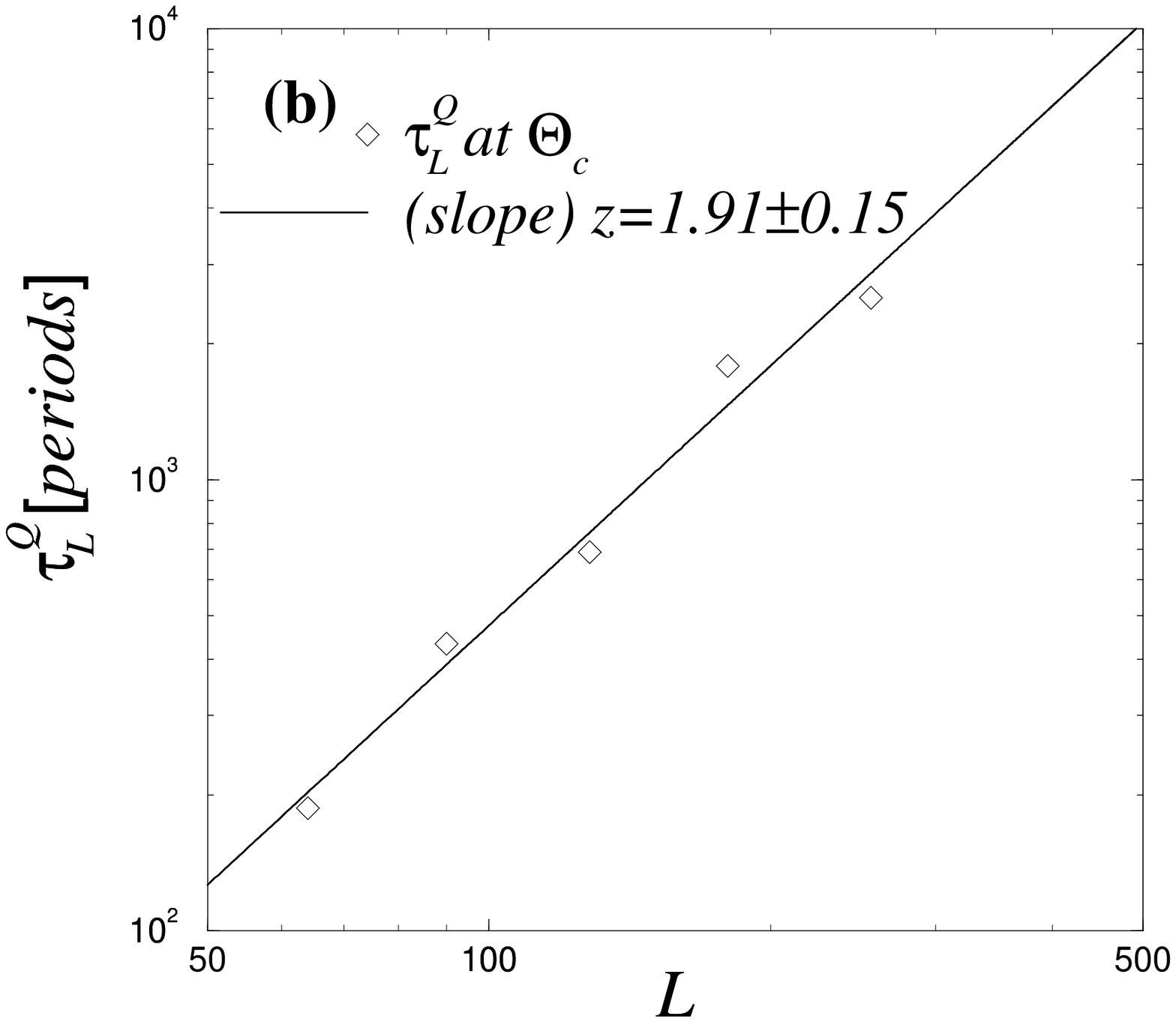}
\caption{Critical slowing down for the order parameter at 
$T$$=$$0.8T_{\rm c}$ and $H_0$$=$$0.3J$ at $\Theta$$=$$\Theta_{\rm c}$. 
(a) The normalized autocorrelation function on lin-log scale for early times. 
The straight lines are fits to exponential decays according to 
Eq.~(\ref{exp_decay}). 
(b) Determining the dynamic exponent, $z$, using a power-law fit to Eq. 
(\protect\ref{power_law}) (log-log plot).}
\label{scale_crit_slow}
\end{figure}

Computing the stationary autocorrelation function given by 
Eq.~(\ref{auto_corr}), we already pointed out that at $\Theta_{\rm c}$
the correlation time increases fast with system size 
[Fig.~\ref{crit_slow}]. 
Correlation times are typically extracted from an exponential decay as 
\begin{equation}
C^{Q}_{L}(n) \propto e^{-n/\tau^{Q}_{L}} \;,
\label{exp_decay}
\end{equation}
and they are expected to be finite for finite systems. The correlation time 
$\tau^{Q}_{L}$ is also well defined in the $L$$\rightarrow$$\infty$ limit
{\em away} from the transition. However, it diverges with $L$ at 
the transition point as
\begin{equation}
\tau^{Q}_{L} \propto L^{z} \;,
\label{power_law}
\end{equation}
where $z$ is the dynamical critical exponent.
For not too late times we had reasonable statistics including the larger
systems (up to $L$$=$$256$) to fit the 
usual exponential decay [Fig.~\ref{scale_crit_slow}(a)]. Then plotting the 
correlation times $\tau^{Q}_{L}$ vs $L$ yields the dynamic 
exponent $z=1.91\pm0.15$, as shown in Fig.~\ref{scale_crit_slow}(b).
This value is within two standard deviations of most estimates for the dynamic 
exponent of the two-dimensional equilibrium Ising model with local dynamics 
\cite{dynamic_z}. 

\subsection{Universality for various temperatures and fields and
cross-over to the strong-field regime}

The underlying ingredient for the spatially extended bistable systems 
exhibiting a DPT is the local metastability (and the corresponding 
characteristic time spent in the metastable ``free-energy well'') in the 
presence of an external field. This, in turn, provides a competition between 
time scales if the system is driven by a periodic field. 
Based on this, we expect that sufficiently large systems (in which many 
droplets contribute to the decay of the metastable phase) exhibit the DPT at 
a half-period $t_{1/2}$ comparable to the metastable lifetime  
$\langle\tau(T,H_0)\rangle$. In other words, we expect the critical
dimensionless half-period to be of order one, $\Theta_{\rm c}\sim{\cal O}(1)$.

To test this expectation,
we performed simulations at $T$$=$$0.8T_c$ for field amplitudes ranging from
$0.3J$ to infinity with system sizes $L$$=$$64$, $90$, and $128$. 
[$H_0$$=$$\infty$ corresponds to the Glauber spin-flip probabilities 
Eq. (\ref{eq:Glauber}) being equal to $0$ (1) depending on whether the spin 
is parallel (anti parallel) to the external field, with no influence from
the configuration of the neighboring spins.] We further 
performed runs at $T$$=$$0.9T_c$, $T$$=$$0.6T_c$, and $T$$=$$0.5T_c$ for 
various field amplitudes and system sizes $L=64$ and $90$. The typical run 
length was $2$$\times$$10^4$ periods. The purpose of these runs was to explore 
the universal nature of the DPT in the multi-droplet regime, and the
crossover to the strong-field regime where the DPT should disappear. 

In the strong-field regime the droplet picture breaks down
since the individual spins are decoupled.
Thus, the metastable phase no longer exists, and the decay of the phase having 
opposite sign to the external field approaches a simple exponential form
(which becomes exact in the $H_0$$\rightarrow$$\infty$ limit).
In the Appendix we show that under these conditions the system magnetization 
always relaxes to a symmetric limit cycle with $Q$$=$$0$ for {\em all} 
frequencies, thus, no DPT can exist.

\begin{figure}
\center
\epsfxsize=5.5cm \epsfysize=5.5cm \epsfbox{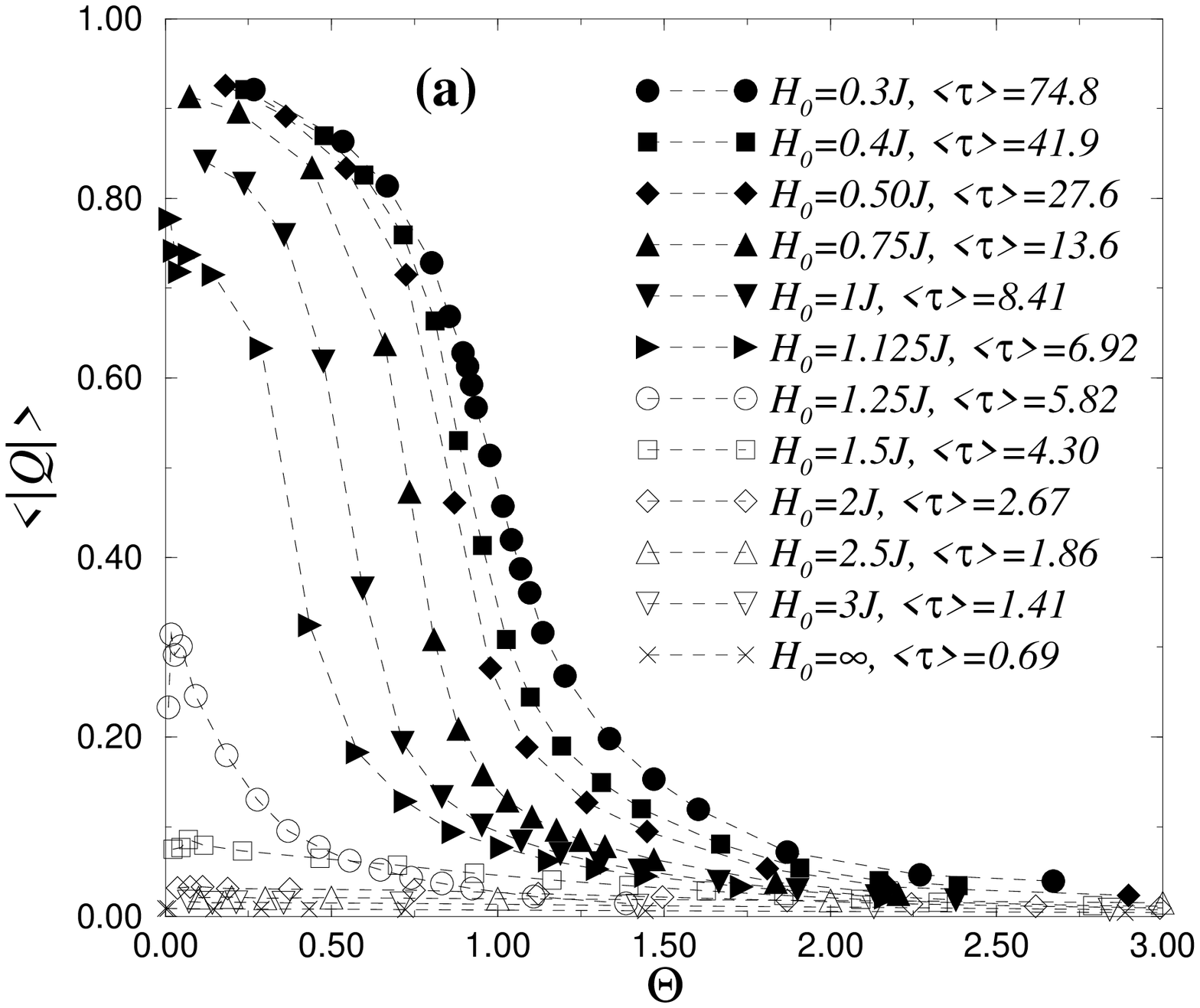}
\epsfxsize=5.5cm \epsfysize=5.5cm \epsfbox{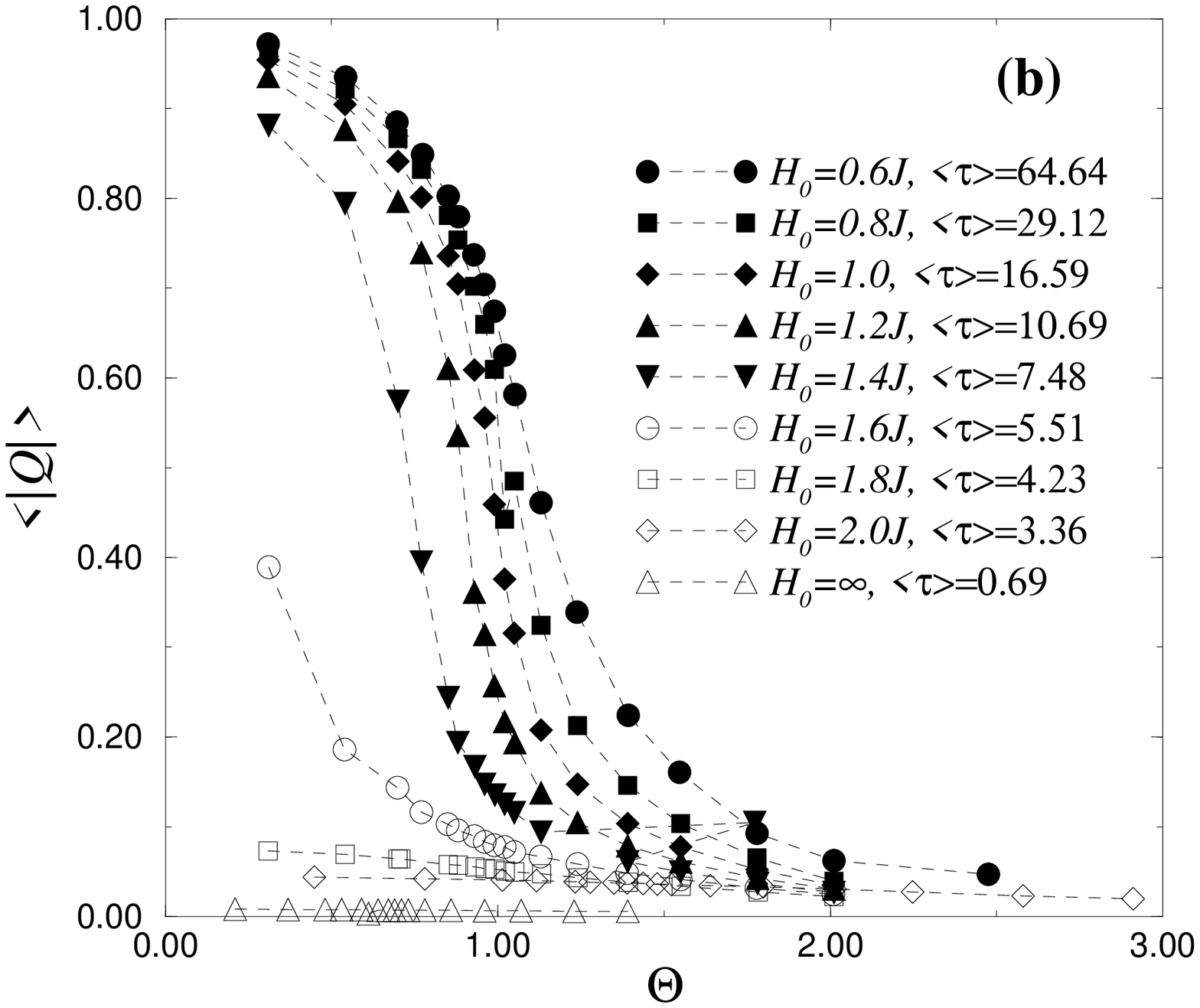}
\epsfxsize=5.5cm \epsfysize=5.5cm \epsfbox{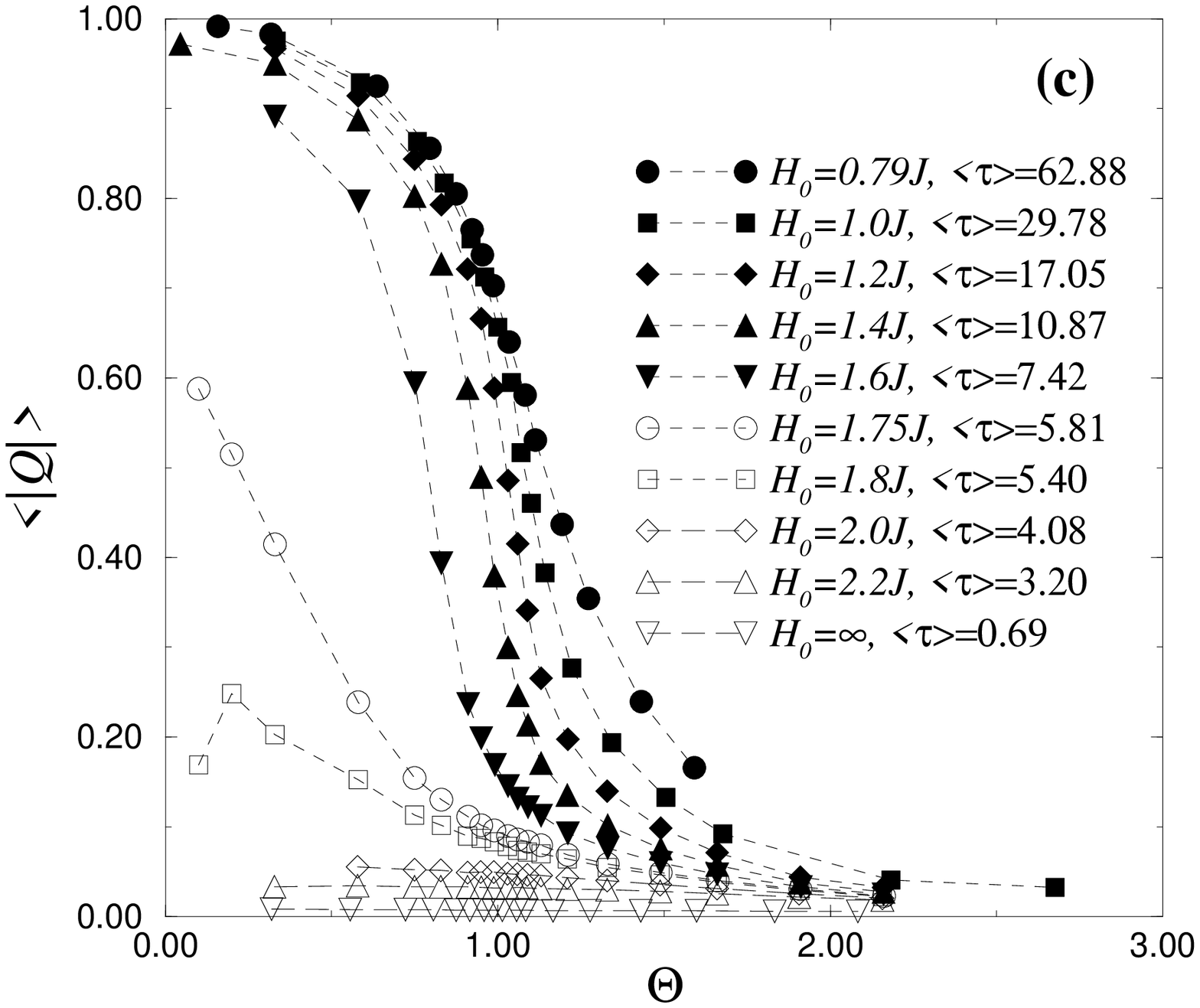}
\caption{DPT in the multi-droplet regime and strong-field cross-over for 
various temperatures and field amplitudes with $L$$=$$64$. Order-parameter 
profiles with filled symbols exhibit a DPT. Corresponding lifetimes in units 
of MCSS are also indicated in the figures. 
(a) $T$$=$$0.8T_c$.
(b) $T$$=$$0.6T_c$.
(c) $T$$=$$0.5T_c$.}
\label{strongfield_cross}
\end{figure}

Figures~\ref{strongfield_cross}(a), (b), and (c) show the order parameter vs
the dimensionless half-period for $L=64$ and  a range of field amplitudes at 
$T$$=$$0.8T_c$, $T$$=$$0.6T_c$, and $T$$=$$0.5T_c$, respectively. The 
typical order-parameter profile where the system exhibits the DPT prevails
up to some temperature dependent cross-over field amplitude $H_{\times}(T)$
[filled symbols in Fig.~\ref{strongfield_cross}]. 
For $H_{0}>H_{\times}(T)$ the underlying decay mechanism belongs to the 
strong-field regime, and correspondingly the DPT disappears as expected.

\begin{figure}
\center
\epsfxsize=5.5cm \epsfysize=5.5cm \epsfbox{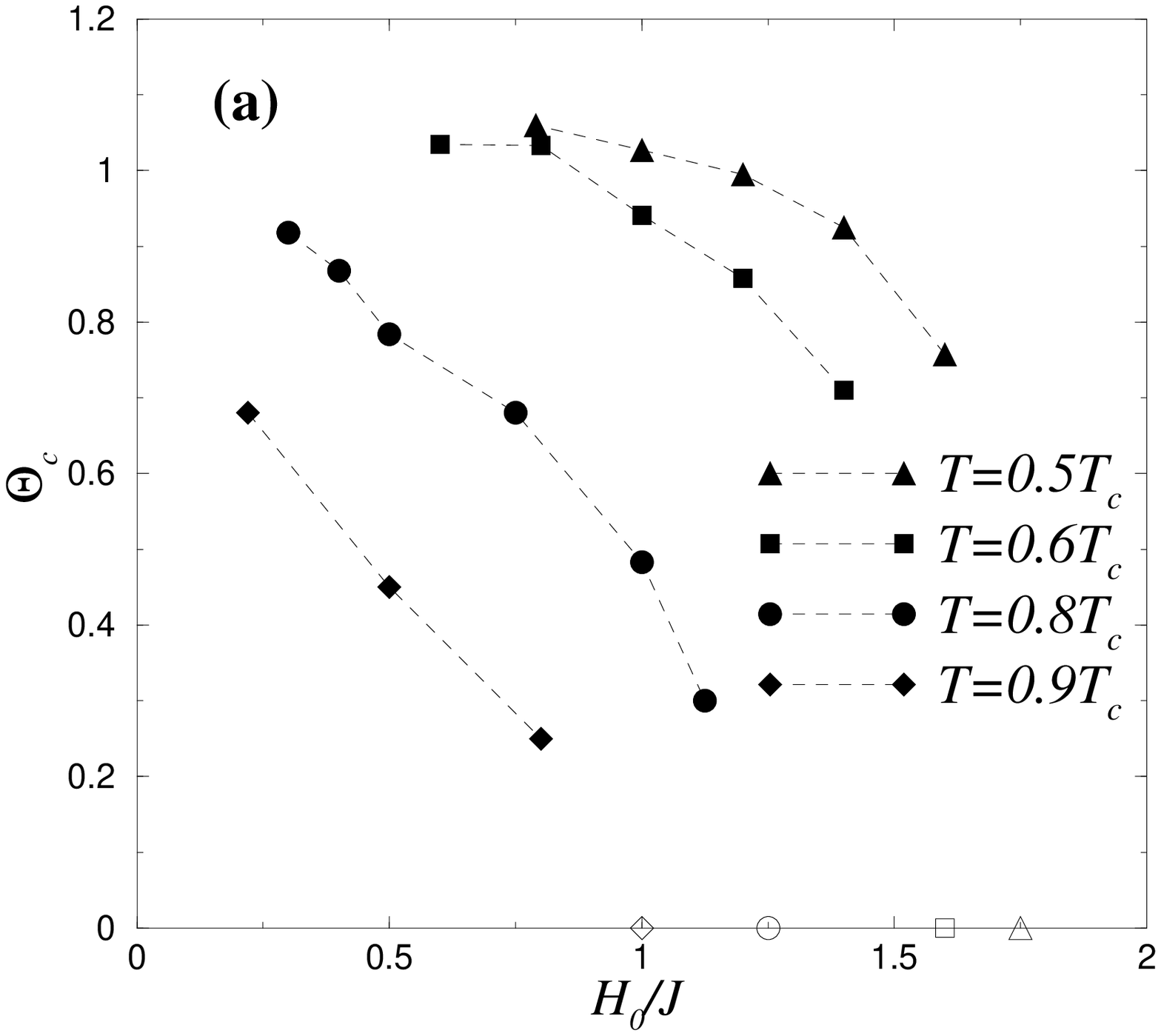}
\epsfxsize=5.5cm \epsfysize=5.5cm \epsfbox{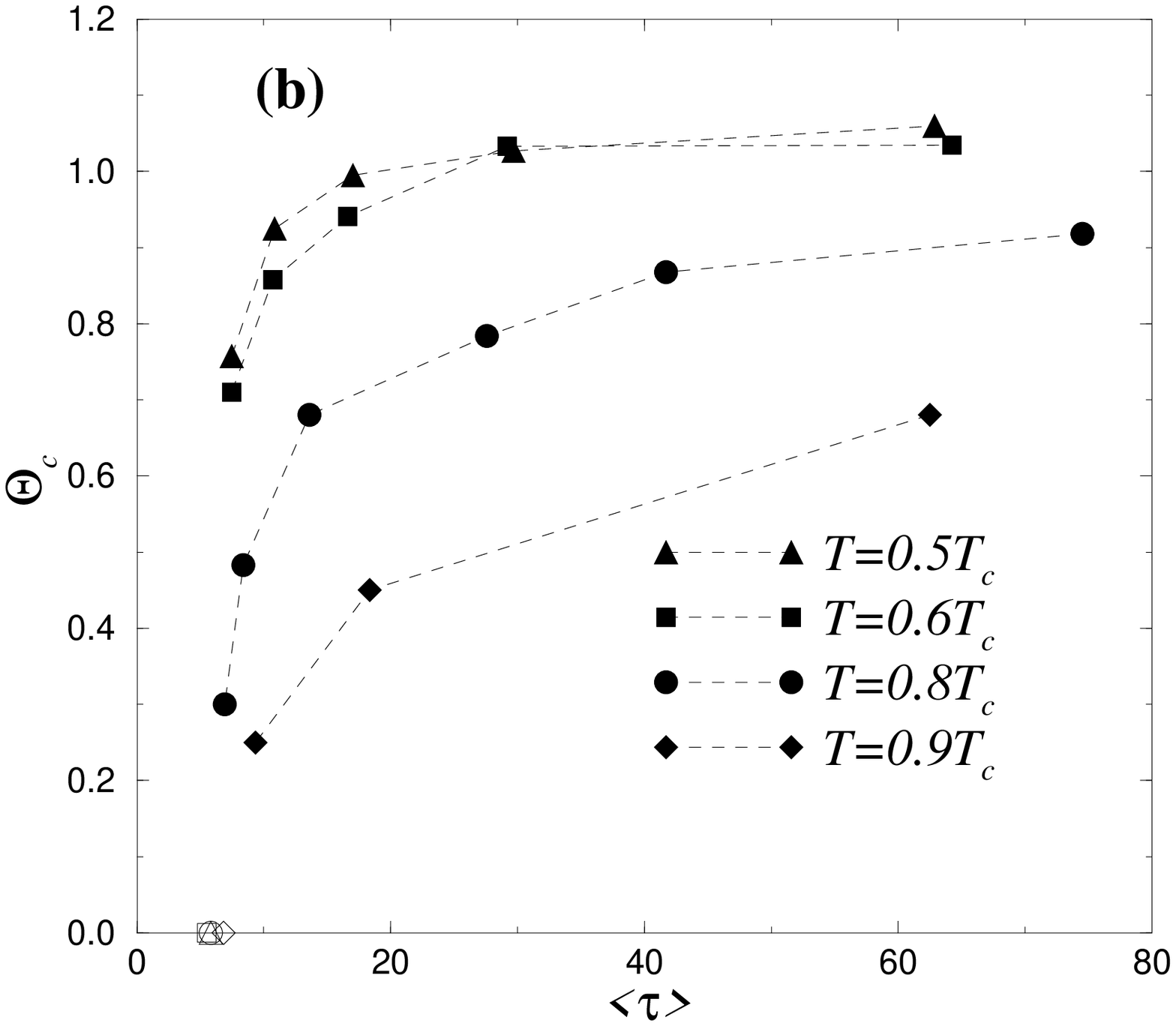}
\epsfxsize=5.5cm \epsfysize=5.5cm \epsfbox{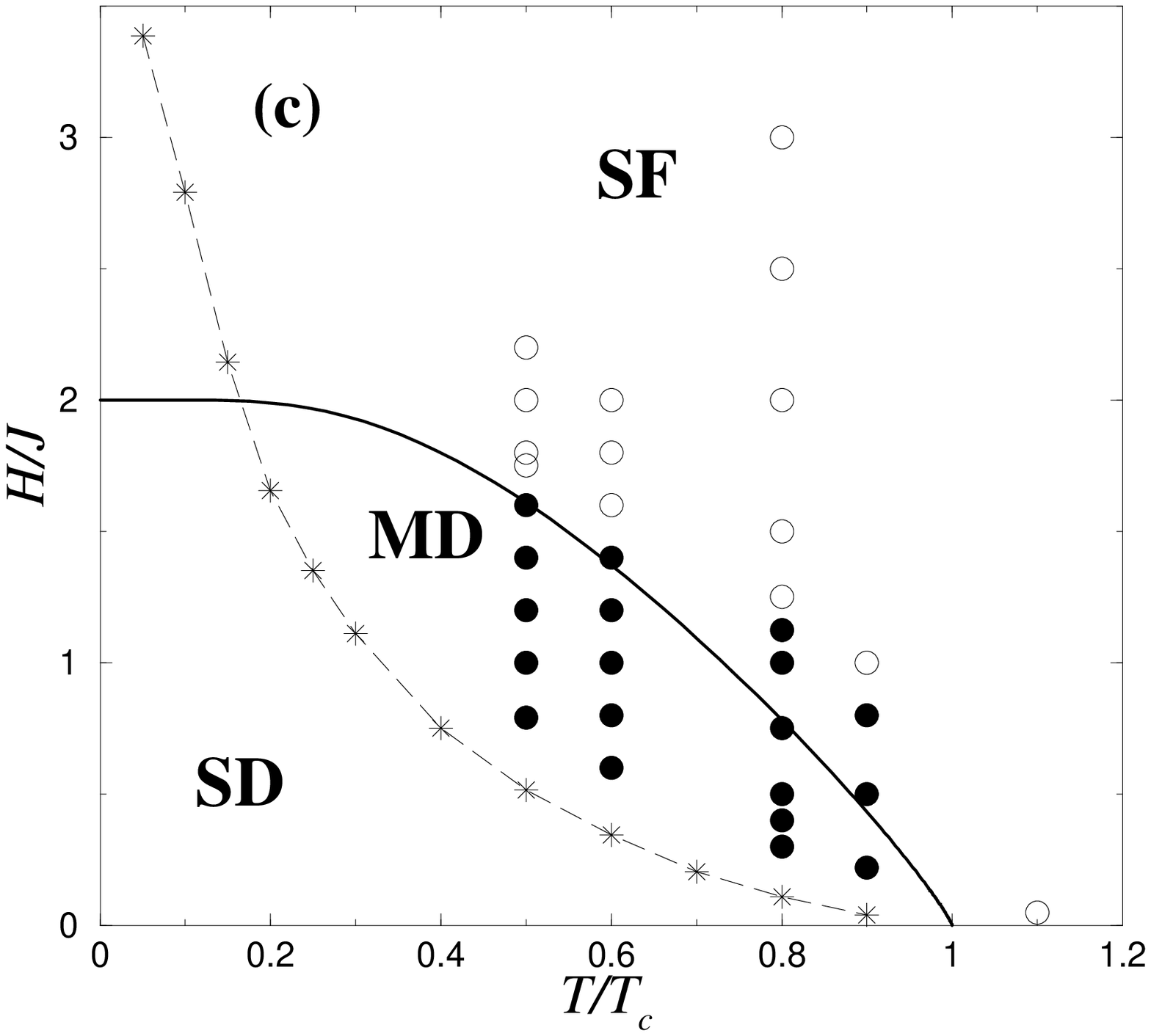}
\caption{Scaled critical half-period in the multi-droplet/DPT regime: (a) as
a function of the external field amplitude $H_0$, (b) as a function of the 
lifetime $\langle\tau\rangle$.
In (a) the corresponding empty symbols for each temperature 
on the horizontal axis represent the points 
which are the MC upper bounds for the cross-over field amplitude, 
$H_{\times}(T)$, beyond which no DPT can be found for any non-zero $\Theta$.
In (b) the corresponding empty symbols are the lower bound for the lifetime
below which no DPT is observed.
(c) Metastable phase diagram for the kinetic Ising model.
The stars connected by a dashed line give a numerical estimate for the 
dynamic spinodal for our
smallest system, $L$$=$$64$, that separates the multi-droplet (MD) from the
single-droplet (SD) regime \protect\cite{switch}. 
The solid line is an analytic estimate for the ``mean-field spinodal'' which 
separates the MD from the strong-field (SF) regime \protect\cite{switch}. 
The circles represent the temperature and field amplitude values at which
we ran our simulations. The filled circles indicate points where the system 
exhibits a DPT, and empty circles represent points where it does not.}
\label{theta_cross}
\end{figure}

One can trace this crossover from the multi-droplet to the strong-field regime
by plotting $\Theta_{\rm c}$ vs $H_{0}/J$ [Fig.~\ref{theta_cross}(a)] and 
vs $\langle\tau\rangle$ [Fig.~\ref{theta_cross}(b)] for fixed temperatures.
Here again we employed the fourth-order cumulant intersection method using
$L$$=$$64$ and $L$$=$$90$ to identify the infinite-system 
transition point, $\Theta_{\rm c}$.
The crossover to the strong-field regime is indicated by the drop in 
$\Theta_{\rm c}$ for large fields (small lifetimes).

\begin{figure}
\center
\epsfxsize=6cm \epsfysize=6cm \epsfbox{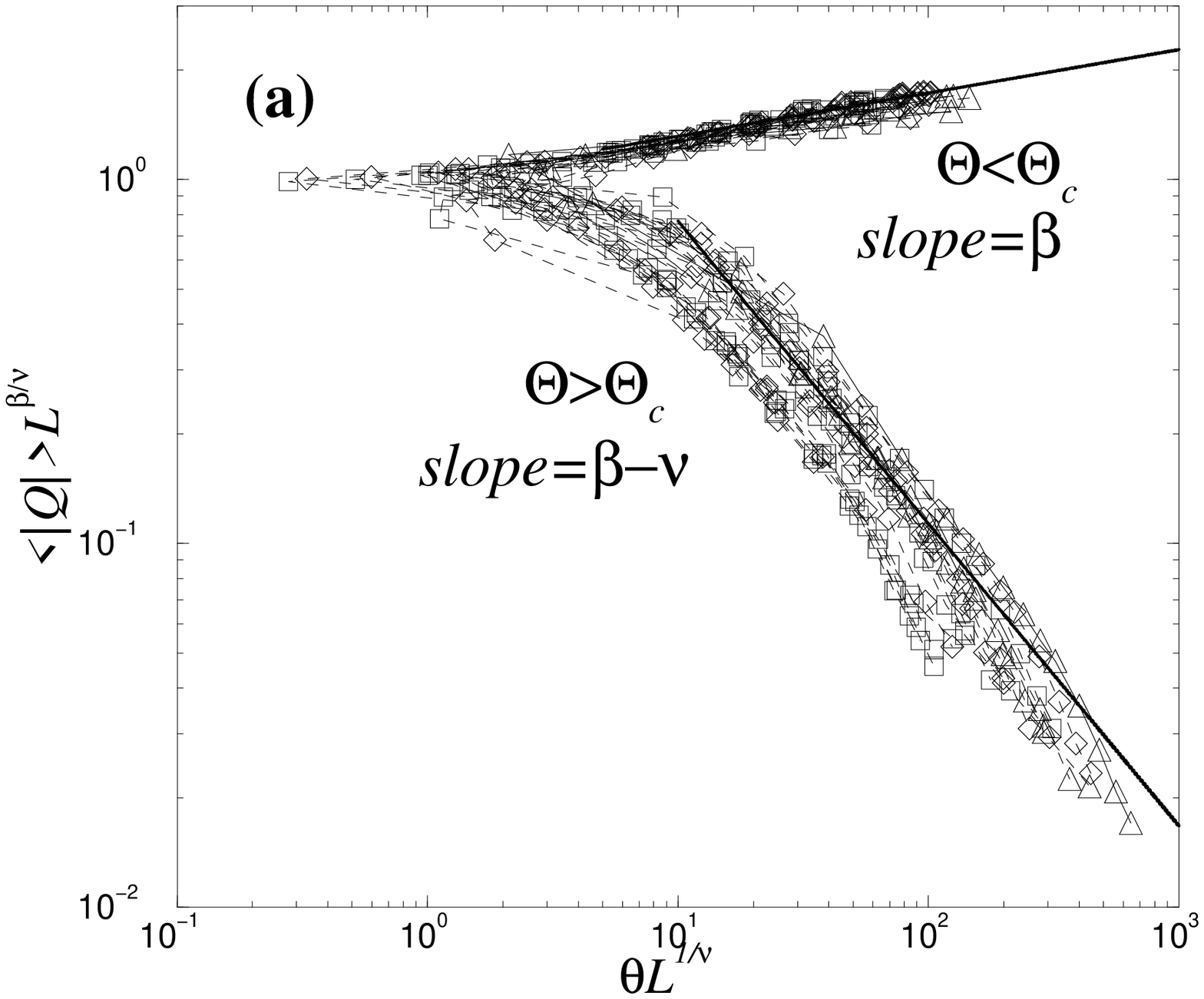}
\epsfxsize=6cm \epsfysize=6cm \epsfbox{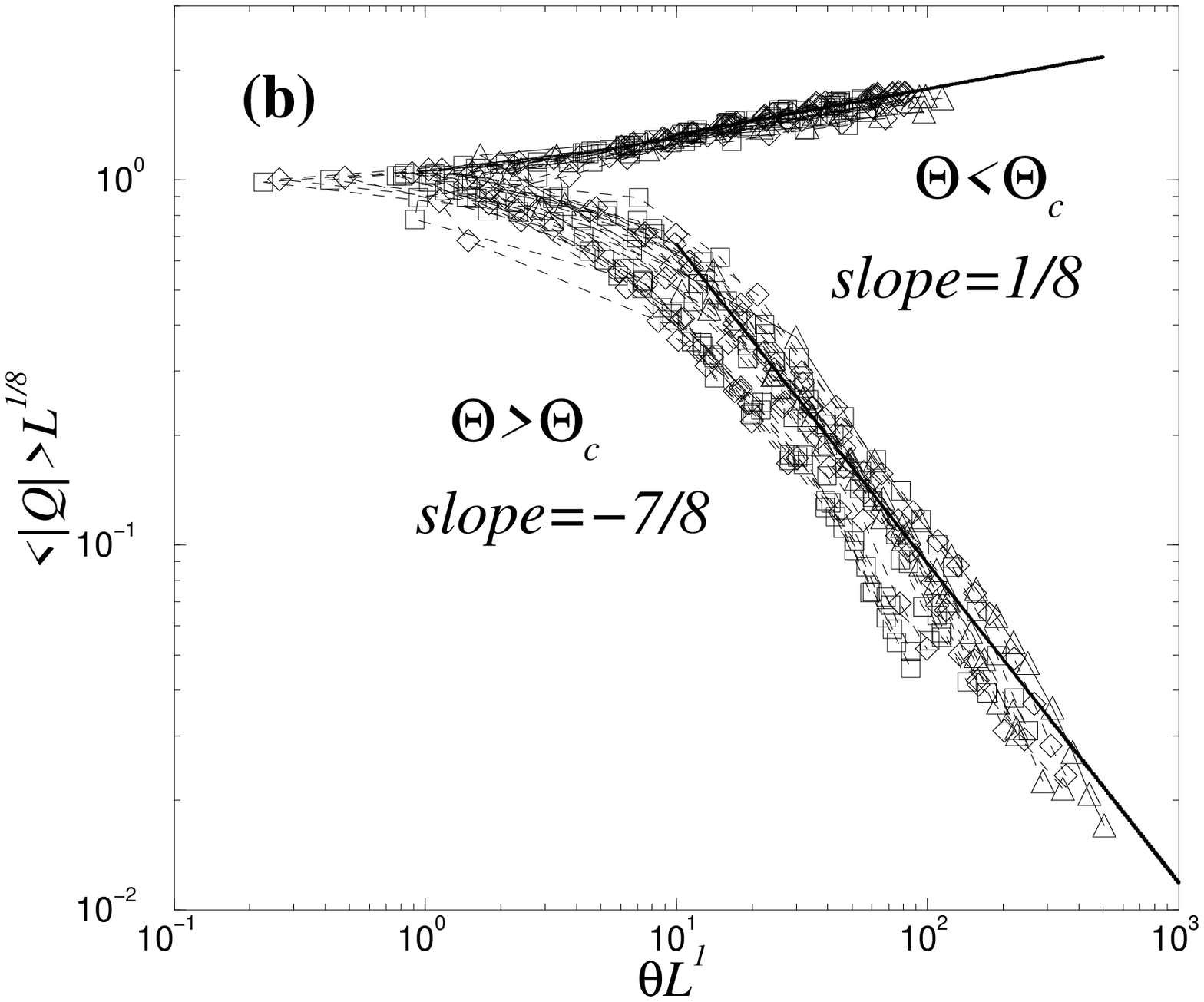}
\caption{Universality of the DPT in the multi-droplet regime. 
(a) Finite-size scaling (full data collapse) for the order parameter 
$\langle|Q|\rangle_L$ (log-log plot).
The figure contains $28$ DPT data sets:
three different temperatures, with four different field amplitudes for each, 
and at least two different system sizes ($L=64,90$, and $128$ at 
$T$$=$$0.8T_c$, and $L=64$ and $90$ at $T$$=$$0.6T_c$ and $T$$=$$0.5T_c$) 
for all these parameters.
The temperature and field values were chosen such that the system is in the 
multi-droplet (MD) regime [filled circles in 
Fig.~\protect\ref{theta_cross}(c)].
We used $\beta/\nu$$=$$1/8$, $\gamma/\nu$$=$$7/4$ 
(two-dimensional equilibrium Ising values), and $\nu$$=$$0.95$, the optimal
value, as described in the Section III.B.
Straight lines represent the asymptotic large-argument 
behaviors of the scaling functions ${\cal F}_{\pm}$ given 
by Eq.~(\ref{full_scaling_Q}). Figure (b) is the same as (a), except that the 
exact Ising exponent $\nu$$=$$1.0 $ is used.}
\label{univ_scale}
\end{figure}

The DPT in the multidroplet regime ($H_{0}$$<$$H_{\times}(T)$) appears to be 
universal, as can be seen from the scaled order-parameter plot in 
Fig.~\ref{univ_scale}. Note that both graphs contain $28$ DPT data sets:
three different temperatures, with four different field amplitudes for each, 
and at least two different system sizes ($L=64,90$, and $128$ at 
$T$$=$$0.8T_c$, and $L=64$ and $90$ at $T$$=$$0.6T_c$ and $T$$=$$0.5T_c$) 
for all these parameters!
While the slopes in the asymptotic scaling regime appear to be the same, the
small parallel shift may be the result of the non-universal critical 
amplitudes at different temperatures and fields, or simply our inaccuracy
in determining $\Theta_{\rm c}(T,H_{0})$, due to the relatively
short runs.

\begin{figure}
\center
\epsfxsize=5.5cm \epsfysize=5.5cm \epsfbox{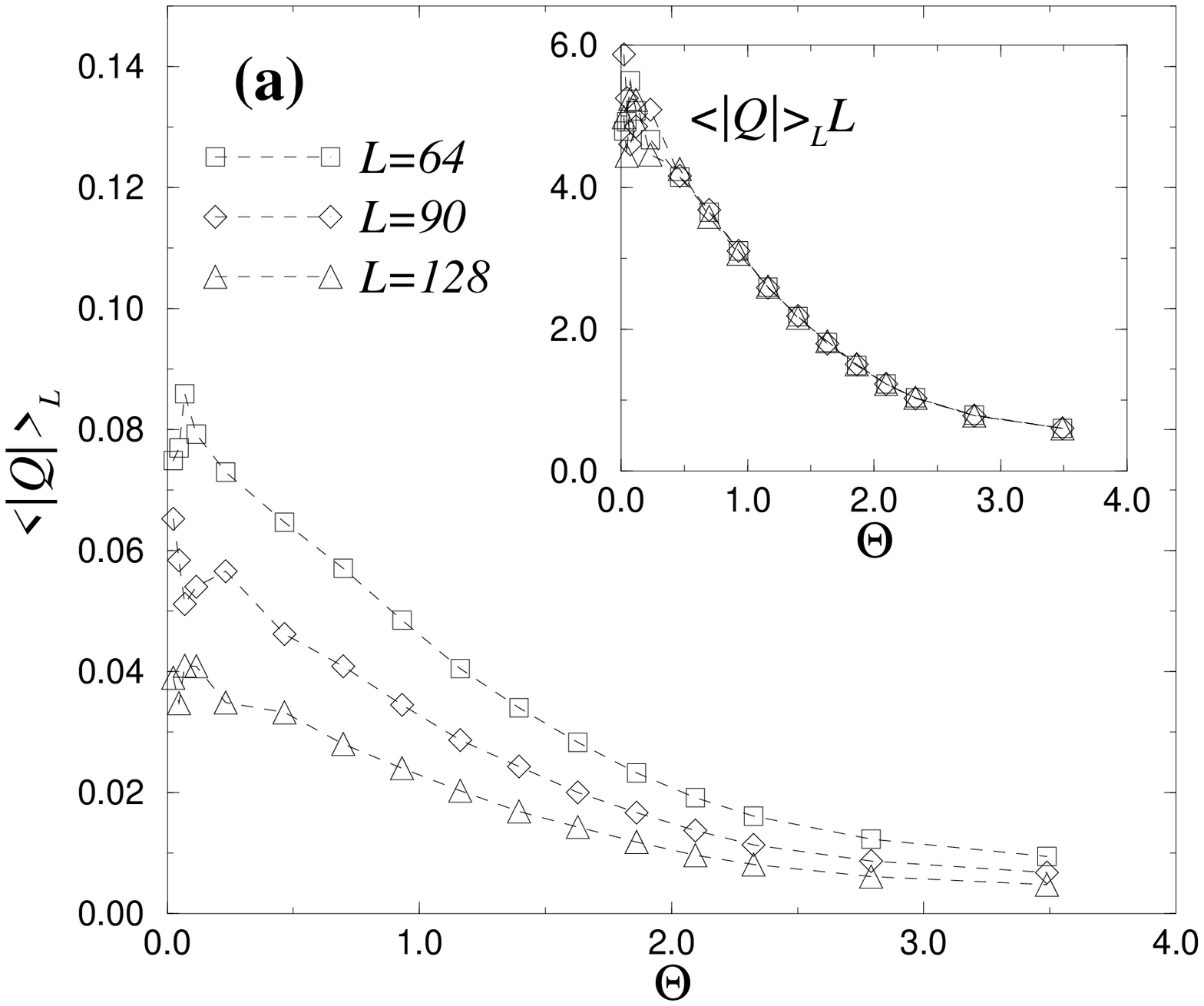}
\epsfxsize=5.5cm \epsfysize=5.5cm \epsfbox{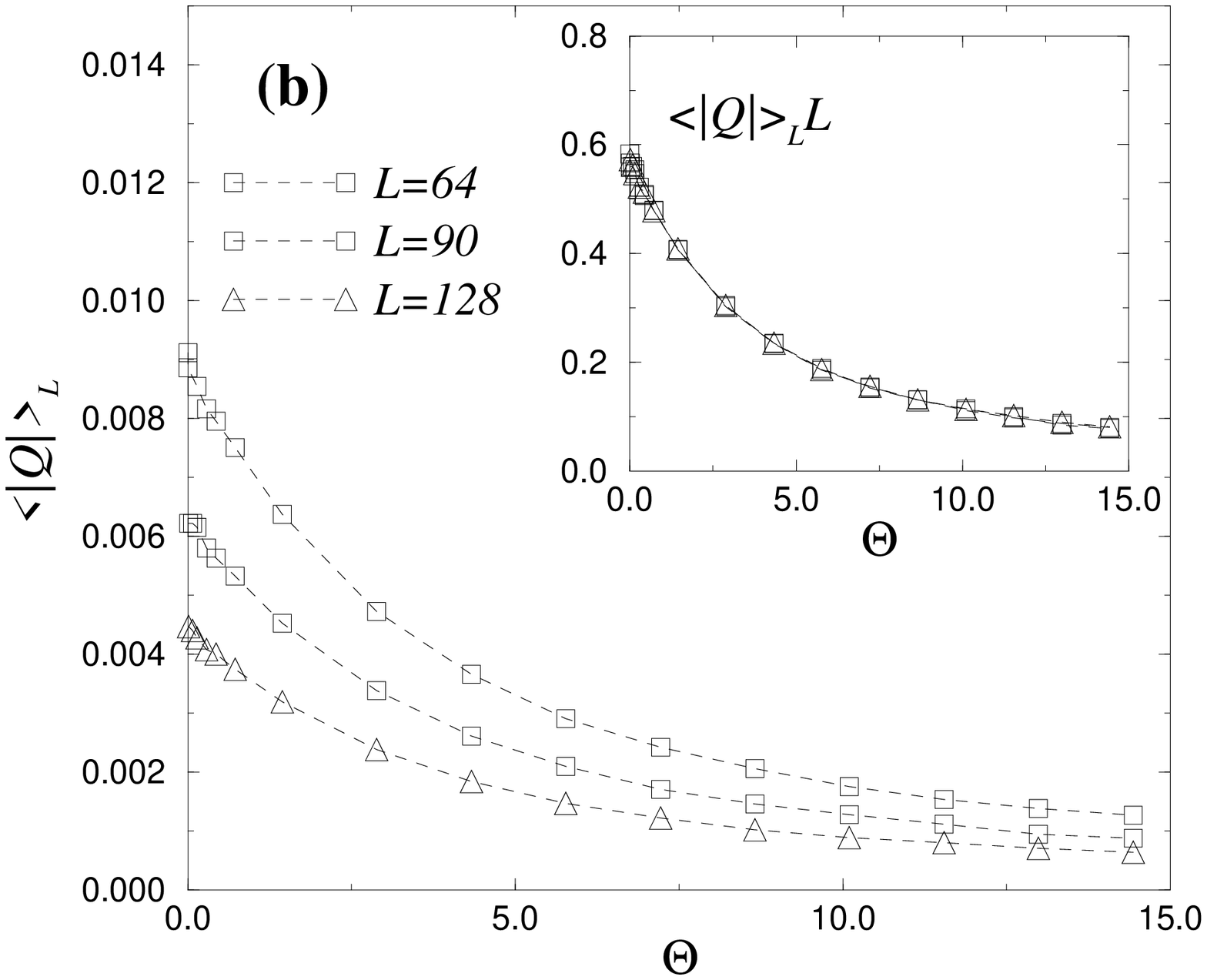}
\epsfxsize=5.5cm \epsfysize=5.5cm \epsfbox{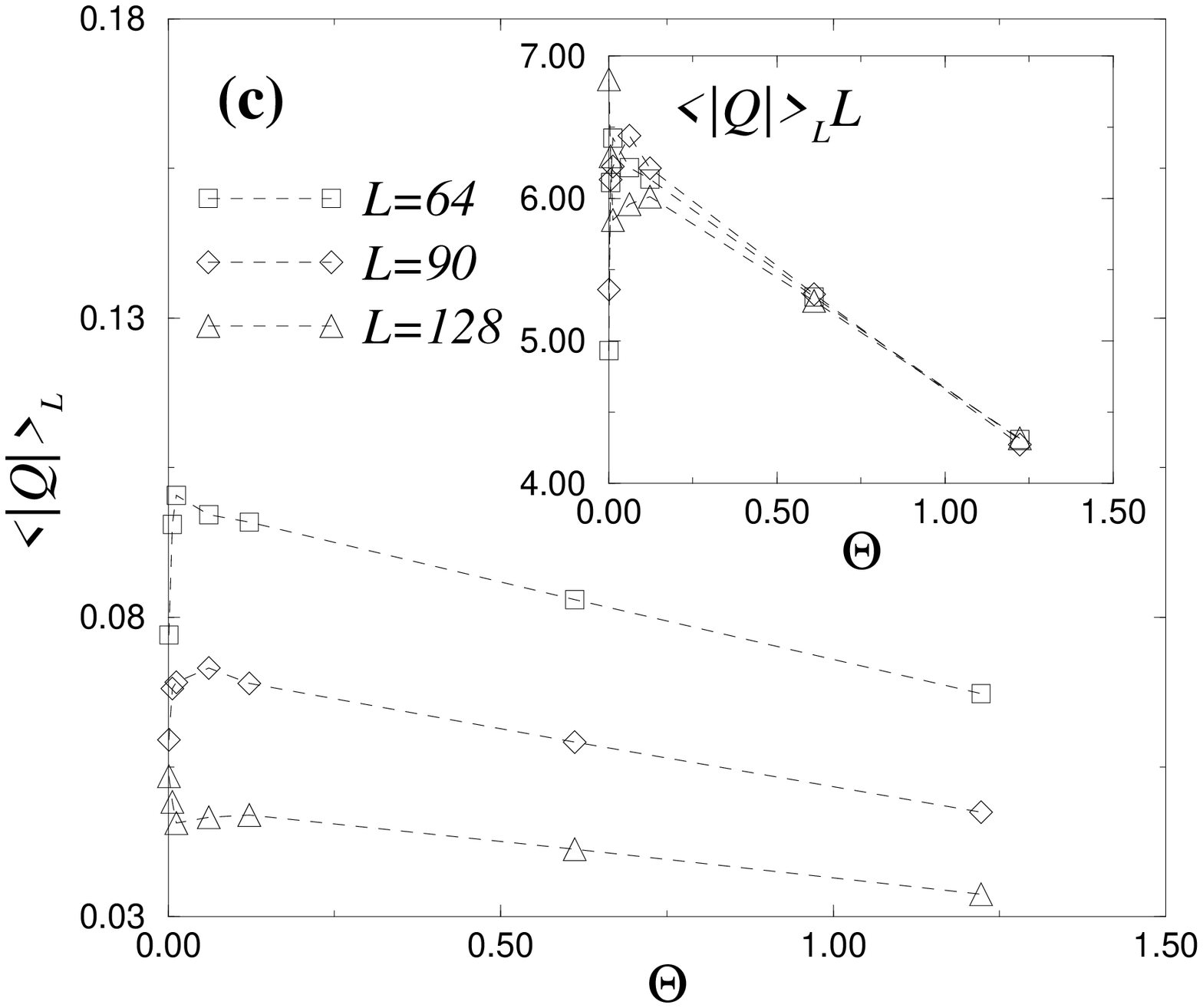}
\caption{Order parameter $\langle|Q|\rangle_{L}$ behavior outside the 
multi-droplet regime.
(a) Strong-field behavior at $T$$=$$0.8T_c$, $H_{0}$$=$$1.5J$ 
($\langle\tau\rangle$$=$$4.3$ MCSS).
(b) Strong-field behavior at $T$$=$$0.8T_c$, $H_{0}$$=$$\infty$ 
($\langle\tau\rangle$$=$$\ln 2$ MCSS).
(c) High-temperature behavior at $T$$=$$1.1T_c$, $H_{0}$$=$$0.05$ 
($\langle\tau\rangle$$=$$81.8$ MCSS).
The insets in all three graphs show data collapse for the scaled order 
parameter $\langle|Q|\rangle_{L}L$.}
\label{no_DPT}
\end{figure}

On the other hand, as expected, the system shows no singularity and 
$\langle |Q|\rangle$$\stackrel{L\rightarrow\infty}{\longrightarrow}$$0$ for 
{\em any} non-zero frequency in the strong-field regime, 
$H_{0}$$>$$H_{\times}(T)$ (see the Appendix). 
Here, the finite-size effects simply reflect the 
central-limit theorem, i.e., $\langle Q^2\rangle$$\sim$${\cal O}(1/L^2)$, or 
$\langle |Q|\rangle$$\sim$${\cal O}(1/L)$. Figures~\ref{no_DPT}(a) and (b)
illustrate this at $T$$=$$0.8T_c$, $H_{0}$$=$$1.5J$ and at $T$$=$$0.8T_c$, 
$H_{0}$$=$$\infty$, respectively.
One also expects that a similar behavior prevails at 
$T$$>$$T_{\rm c}$ for {\em any} field amplitude, i.e., 
$H_{\times}(T)$ vanishes for $T$$>$$T_{\rm c}$.
We checked this for $T$$=$$1.1T_c$ and $H_{0}$$=$$0.05J$, and
the results confirm the $\langle |Q|\rangle$$\sim$${\cal O}(1/L)$ scaling
of the order parameter for all frequencies, showing none of the characteristic 
finite-size effects of a DPT [Fig.~\ref{no_DPT}(c)].
This result is in distinct disagreement with older work such as 
Refs.~\cite{ACHA95,ACHA94}, which report observations of the DPT at 
temperatures considerably above $T_{\rm c}$. 
For smaller system sizes, however, one may observe nontrivial resonance 
\cite{NEDA}.

Due to the expected complications of a divergent {\em equilibrium} correlation
length, we did not perform simulations at $T_{\rm c}$. However, we conjecture
that $H_{\times}(T)$ vanishes at $T_{\rm c}$. We expect this conjecture to be 
extremely difficult to prove or disprove numerically, due to very large and 
possibly complicated finite-size effects.

\section{Conclusions and Outlook}

In this paper we have studied the hysteretic response of a spatially extended
bistable system exhibiting a DPT. Our model system is
the two-dimensional kinetic Ising ferromagnet below its equilibrium critical 
temperature subject to a periodic square-wave applied field. The results 
indicate that for field amplitudes and temperatures such that the metastable 
phase 
decays via the multi-droplet mechanism, the system undergoes a continuous
dynamic phase transition when the half-period of the field, $t_{1/2}$, is 
comparable to the metastable lifetime, $\langle\tau(T,H_0)\rangle$. Thus,
the critical value $\Theta_{\rm c}$ of the dimensionless half-period defined
in Eq.~(\ref{eq:Theta}) is of order unity. As $\Theta$ is increased beyond 
$\Theta_{\rm c}$, the order parameter $\langle|Q|\rangle$ (the expectation
value of the norm of the period-averaged magnetization) vanishes
[Fig.~\ref{Q_raw}(a)], displaying singular behavior at the critical point, 
as shown in Fig.~\ref{Q_raw}(b) and (c).

The characteristic finite-size effects in the order parameter and its 
fluctuations indicate that there is a divergent correlation length associated
with the transition. We used standard finite-size scaling techniques adopted
from the theory of equilibrium phase transitions. We estimated 
$\Theta_{\rm c}$ and the critical exponents $\beta$, $\gamma$, and $\nu$
from relatively high precision data for system sizes between $L=64$ and $512$ 
at $T$$=$$0.8T_{\rm c}$ and $H_{0}$$=$$0.3J$.  
Our best estimates are $\beta/\nu$$=$$0.126\pm 0.005$, 
$\gamma/\nu$$=$$1.74\pm 0.05$, and $\nu$$=$$0.95\pm 0.15$. These values
agree within statistical errors with those previously obtained with a 
sinusoidally
oscillating field \cite{SIDES99,SIDES98}, providing strong evidence that the 
shape of the field oscillation does not affect the universal aspects of the 
DPT. Observing the stationary autocorrelation function we also saw that at the 
transition point the system exhibits critical slowing down governed by the  
dynamic exponent $z$$=$$1.91\pm0.15$. This is also very close to the 
corresponding exponent $z$$=$$2.12(5)$, measured in standard two-dimensional 
Ising simulations with local dynamics \cite{dynamic_z}.
Our best values for the exponent ratios $\beta/\nu$ and $\gamma/\nu$ are
given with relatively high confidence, while for $\nu$ it is rather poor. 
In this sense tracking down the exponent $\nu$ and obtaining an accurate 
estimate for it remains elusive. Note, however, that we could only
rely on the standard (single spin-flip) MC algorithm, since we had to preserve 
the underlying dynamics. 
Using more sophisticated algorithms to avoid critical 
slowing down as seen in Figs.~\ref{crit_slow} and \ref{scale_crit_slow}, 
would require an underlying ``Hamiltonian'' for the 
corresponding {\em local} order parameter $\{Q_i\}$, which is not yet known.
While in a coarse-grained/universal sense a $\phi^4$ Hamiltonian is supported 
by our data, it does not point to any one particular microscopic Hamiltonian 
for the microscopic order parameter $\{Q_i\}$. 
However, a $\phi^4$ coarse-grained Hamiltonian for $\{Q_i\}$ has been 
recently derived starting from the time-dependent Ginzburg-Landau equation 
for the magnetization\cite{Fuji}.

Of the known universality classes, our exponent estimates for the DPT are 
closest (and within the statistical errors) to those of the the 
two-dimensional equilibrium Ising model:
$\beta/\nu$$=$$1/8$$=$$0.125$, $\gamma/\nu$$=$$7/4$$=$$1.75$, and
$\nu$$=$$1$. Consequently, our measured exponent ratios satisfy the 
hyperscaling relation
\begin{equation}
2(\beta/\nu)+\gamma/\nu = 1.99\pm 0.05 \approx d \;,
\end{equation}
where $d$$=$$2$ is the spatial dimension \cite{endnote2}. 
Further, the fixed-point value of the 
fourth-order cumulant, $U^*$$=$$0.611$$\pm$$0.003$, is also extremely close to 
that of the Ising model, $U^*$$=$$0.6106901(5)$ \cite{BLOTE}. These findings 
provide conclusive evidence that the DPT indeed corresponds to a non-trivial 
fixed point. 
We tested the full data collapse for the scaled order parameter and its 
variance, as shown in Fig.~\ref{full_collapse}, and it confirmed the existence
of the universal scaling functions given by Eqs. (\ref{full_scaling_Q}) and 
(\ref{full_scaling_XQ}).
Also, at the critical frequency, the order-parameter distributions follow
finite-size scaling predictions, Eq.~(\ref{scaling_PQ}), as shown in 
Fig.~\ref{scaled_hist}.
More surprisingly, the critical 
DPT order-parameter distributions fall on that of the two-dimensional 
equilibrium Ising model at the critical temperature (except for stronger 
corrections to scaling), {\em without}  any additional fitting of the 
underlying microscopic length scale.

While our finite-size scaling data clearly indicate the existence of a 
divergent length scale, we did not measure the correlation length for
the {\em local} order parameter directly. Future studies may include extracting
the correlation length, $\xi_Q$, from the $\langle Q_i Q_j\rangle$ 
correlations (or from the corresponding structure factor). This approach 
would also provide another way to measure the exponent $\nu$ by plotting 
$\xi_Q$ vs $\theta$ in the critical regime for large systems, and assuming  
$\xi_Q$$\sim$$\theta^{-\nu}$.

We also studied the universal aspects of the DPT at other temperatures and 
fields. Shorter runs at $T$$=$$0.8T_{\rm c}$, $T$$=$$0.6T_{\rm c}$, and 
$T$$=$$0.5T_{\rm c}$ for field amplitudes $H_{0}$$<$$H_{\times}(T)$
also confirmed scaling and the universality of the DPT 
[Fig.~\ref{univ_scale}]. The condition for the field amplitude implies
that the system only exhibits a DPT in the multi-droplet regime.
For $H_{0}$$>$$H_{\times}(T)$ strong-field behavior governs the 
decay of the magnetization, and the DPT disappears, as indicated by 
Fig.~\ref{theta_cross} and Fig.~\ref{no_DPT}(a),(b). We also found that the 
high-temperature phase is qualitatively similar to the strong-field regime in 
that there is no sign of a DPT for $T$$>$$T_{\rm c}$ [Fig.~\ref{no_DPT}(c)].

One may ask how general the phenomenon of a DPT is in spatially extended 
bistable systems, subject to a periodic applied ``field'' which drives the 
system between its metastable and stable ``wells.'' It is possible that 
having up-down symmetry of the period-averaged ``magnetization'' is sufficient
for possessing a Hamiltonian at the coarse-grained level, even if the 
system is driven far away from equilibrium and is microscopically
irreversible \cite{GRINSTEIN}.
Future research can address this question by studying other systems (not 
necessarily ferromagnets) that exhibit hysteresis.

\acknowledgments
We thank S.~W.\ Sides, S.~J.\ Mitchell, and G.\ Brown for stimulating 
discussions. We would like to thank W. Janke for providing us with  
data from Ref.~\cite{Janke} for comparison of the critical order-parameter 
distributions.
We acknowledge support by the US Department of Energy through the former 
Supercomputer Computations Research Institute, by the Center for Materials 
Research and Technology at Florida State University, and by the US National
Science Foundation through Grants No. DMR-9634873, DMR-9871455,
and DMR-9981815.
This research also used resources of the National Energy Research Scientific 
Computing Center, which is supported by the Office of Science of the U.S.
Department of Energy under Contract No.\ DE-AC03-76SF00098.

\appendix
\section*{Low-frequency and Strong-field Mean-field 
Approximation}

For simplicity, in the following we assume that the magnetization decays from
$m$$=$$+1$ to $m$$=$$-1$ after a {\em single} field reversal 
($H_0$$\rightarrow$$-H_0$). 
This is a good approximation below the equilibrium critical temperature 
($m_{\rm sp}$$\approx$$1$), and for any temperature when
$H_0$$\rightarrow$$\infty$.
Further, we assume that the volume fraction of meta- or unstable spins
follows a simple monotonic decay, $\hat{\varphi}(t)$.

In terms of the volume fraction of {\em positive} spins, $\phi(t)$, 
the magnetization can be written as 
\begin{equation}
m(t) = 2\phi(t) - 1 \;.
\end{equation}
Subject to a square-wave field,
\begin{equation}
H(t) = \left\{ \begin{array}{ll}
-H_0 & 0 \leq t <t_{1/2} \\
+H_0 & t_{1/2} \leq t <2t_{1/2}
\end{array} \right. \;,
\end{equation} 
in the first (second) half-period the volume fraction of the 
positive (negative) spins decays according to $\hat{\varphi}(t)$.
Thus, in each period (measuring time $t$ from the beginning of the period)
\begin{equation}
\phi(t) \approx \left\{ \begin{array}{ll}
\phi(0)\hat{\varphi}(t) & 0\leq t \leq t_{1/2} \\
1-\left[1-\phi(t_{1/2})\right ]\hat{\varphi}(t-t_{1/2}) & 
t_{1/2} \leq t \leq 2t_{1/2}
\end{array} \right. \;.
\end{equation}
Using this approximation, one directly obtains a linear mapping  
\begin{equation}
\phi_{n+1} = 1-[1-\phi_n\hat{\varphi}(t_{1/2})]
 \hat{\varphi}(t_{1/2}) \;,
\end{equation}
where $\phi_n$$\equiv$$\phi(2nt_{1/2})$ is the volume fraction of the 
positive spins at the beginning of the $n$th period, $n=0,1,2,\ldots$.
The stationary value of this quantity is
\begin{equation}
\phi^* = \lim_{n\rightarrow\infty}\phi_n = 
\frac{1}{1+\hat{\varphi}(t_{1/2})} \;.
\end{equation}
Consequently, the magnetization reaches a stationary limit cycle
\begin{equation}
m(t) \approx \left\{ \begin{array}{ll}
\frac{2}{1+\hat{\varphi}(t_{1/2})}\hat{\varphi}(t) - 1 & 
0 \leq t \leq t_{1/2} \\
1 - \frac{2}{1+\hat{\varphi}(t_{1/2})}\hat{\varphi}(t-t_{1/2}) &
t_{1/2} \leq t \leq 2t_{1/2}
\end{array} \right. \;.
\end{equation}
In this limit cycle the magnetization oscillates about zero,
\begin{equation}
m(0) = - m(t_{1/2}) =
\frac{1-\hat{\varphi}(t_{1/2})}{1+\hat{\varphi}(t_{1/2})} \;
\end{equation}
and the symmetry of the magnetization, $m(t\pm t_{1/2})$$=$$-m(t)$ implies
\begin{equation}
Q = \frac{1}{2 t_{1/2}} \oint m(t) dt = 0\;.
\end{equation}
This corresponds to the symmetric (dynamically disordered) phase. Note that
this symmetric phase is always reached when $\hat{\varphi}$ decreases
monotonically from unity  at $t$$=$$0$ to zero as $t$$\rightarrow$$\infty$

In the multi-droplet regime, the volume fraction of the metastable phase decays
according to Avrami's law \cite{Avrami,RAMOS99}. 
In the low-frequency limit, $t_{1/2}$$\gg$$\langle\tau\rangle$ 
($\Theta$$\gg$$1$), each half-period 
almost always contains a {\em complete} metastable decay 
[Fig.~\ref{m_series}(a)]. 
Avrami's law for the metastable volume fraction in each half-period can 
then be directly applied using 
$\hat{\varphi}(t)$$=$%
$\varphi_{\rm ms}(t)$$\approx$$e^{-(\ln2)t^3/\langle\tau\rangle^3}$ 
\cite{Avrami,RAMOS99}. This 
functional form for $\varphi_{\rm ms}(t)$ breaks down when $t_{1/2}$ becomes 
comparable to $\langle\tau\rangle$ ($\Theta$$\approx$$1$), thus this simple 
mean-field approximation cannot predict any instability related to the DPT.

In the $H_0$$\rightarrow$$\infty$ limit the individual spins become  
decoupled. Then one can obtain 
$\hat{\varphi}(t)$$=$$e^{-(\ln2)t/\langle\tau\rangle}$ 
which is exact for {\em all} frequencies.
Further, this exponential decay is a good approximation everywhere in the 
strong-field regime, thus, no DPT can exist there.


\begin{references}
\bibitem{endnote1} 
The difference between the order-parameter time series in 
Fig.~\ref{Q_series} (strongly fluctuating trace) and Fig.~\ref{series_hist}(a)
is due to that the latter is taken at the infinite-system transition 
point ($\Theta$$=$$\Theta_{\rm c}$$=$$0.918$), while the former is taken at the
corresponding finite-system ``critical point'' 
($\Theta$$=$$\Theta_{\rm c}(L)$$=$$0.98$), where the fluctuations are largest.
\end{references}
\end{document}